\documentclass[a4paper,11pt]{article}
\pdfoutput=1
\usepackage{graphicx}
\usepackage{epsf,amsmath,bbold,amsfonts,stmaryrd}
\usepackage{amsmath,amssymb,amsthm,mathrsfs,amsfonts,dsfont}
\usepackage[utf8]{inputenc}
\usepackage{mathrsfs}
\usepackage{appendix}
\usepackage{amssymb}
\usepackage{float}
\usepackage{color}
\usepackage{cite}
\usepackage{hyperref}
\hypersetup{pageanchor=false}
\usepackage{indentfirst}
\usepackage{url}
\usepackage{float}
\usepackage{caption}
\usepackage[numbers,square,comma,sort&compress,merge]{natbib}
\usepackage{graphicx}
\usepackage{animate}
\usepackage{subfigure}

\def\be{\begin{equation}}
\def\ee{\end{equation}}

\def\bea{\begin{eqnarray}}
\def\eea{\end{eqnarray}}

\def\ba{\begin{array}}
\def\ea{\end{array}}

\def\bc{\begin{center}}
\def\ec{\end{center}}

\def\bl{\begin{flushleft}}
\def\el{\end{flushleft}}

\def\br{\begin{flushright}}
\def\er{\end{flushright}}

\def\bi{\begin{itemize}}
\def\ei{\end{itemize}}

\def\bt{\begin{tabular}}
\def\et{\end{tabular}}

\numberwithin{equation}{section}

\begin{document}

\title{\textbf{Bending the Bruhat-Tits Tree I \\
Tensor Network and Emergent Einstein Equations }}
\author{Lin Chen, Xirong Liu \footnote{Chen and Liu  are co- first authors of the manuscript.}, and Ling-Yan Hung\\
   $^1$State Key Laboratory of Surface Physics, \\
    Fudan University, \\
    200433 Shanghai, China\\
    $^2$Shanghai Qi Zhi Institute, \\
    41st Floor, AI Tower, No. 701 Yunjin Road,  \\
    Xuhui District, Shanghai, 200232, China\\
    $^3$Department of Physics and Center for Field Theory and Particle Physics, \\
    Fudan University, \\
    200433 Shanghai, China\\
    $^4$Institute for Nanoelectronic devices and Quantum computing, \\
    Fudan University, \\
    200433 Shanghai , China\\}
\date{\today}

\date{}

\maketitle

\vspace{-10mm}

\vspace{8mm}

\begin{abstract}
As an extended companion paper to \cite{paper1}, we elaborate in detail how the tensor network construction of a p-adic CFT encodes geometric information of a dual geometry even as we deform the CFT away from the fixed point by finding a way to assign distances to the tensor network. In fact we demonstrate that a unique (up to normalizations) emergent graph Einstein equation is satisfied by the geometric data encoded in the tensor network, and the graph Einstein tensor automatically recovers the known proposal in the mathematics literature, at least perturbatively order by order in the deformation away from the {\it pure Bruhat-Tits Tree} geometry dual to pure CFTs. Once the dust settles, it becomes apparent that the assigned distance indeed corresponds to some Fisher metric between quantum states encoding expectation values of bulk fields {\it in one higher dimension}.
This is perhaps a first quantitative demonstration that a concrete Einstein equation can be extracted directly from the tensor network, albeit in the simplified setting of the p-adic AdS/CFT.

\end{abstract}
\baselineskip 18pt

\thispagestyle{empty}
\newpage

\tableofcontents

\section{Introduction}

The Ryu-Takanayagi formula provides very deep insight into the underlying physics of the AdS/CFT correspondence \cite{Ryu:2006bv}. It suggests that the gravity dual is encoding the entanglement structure of the CFT in a geometrical way \cite{VanRaamsdonk:2009ar}. This is strongly reminiscent of tensor networks, which captures the entanglement of a given many body state it approximates through a prudent choice of graph the network covers.  This led to suggestions that the tensor network is perhaps the microscopic explanation of the AdS/CFT correspondence \cite{Swingle:2009bg}.

Many models have been proposed since, capturing different aspects of the AdS/CFT, in addition to the RT formula that first inspired the analogy.
The perfect tensor and the random tensor networks \cite{Pastawski:2015qua, Hayden:2016cfa} capture the error correcting properties \cite{Almheiri:2014lwa}, and explicitly realized the notion of entanglement wedge reconstruction.
More recently there are different algorithms developed to grow a bulk tensor network. See for example \cite{Bao:2019fpq,Lin:2020thc, Gan:2017nyt,Chen:2018ywy}.

The tensor network also makes ready connection with notions of complexities \cite{Susskind:2014rva,Brown:2015lvg,Chapman:2016hwi, Carmi:2016wjl,Jefferson:2017sdb,Caputa:2017yrh,Czech:2017ryf, Bhattacharyya:2019kvj}, allowing visualisation of complexity growth of the wavefunction simply through the counting of tensors, giving extra support to the idea of volume/action of the AdS bulk being a measure of complexity.

The tensor network also gives an intuitive picture to the island formula \cite{Hartman:2020khs}, the latest breakthrough that potentially explains the black hole information paradox \cite{Penington:2019npb,Almheiri:2019hni}.

The list of successes give strong support that the tensor network picture is capturing some important essence of the AdS/CFT correspondence.
One very important challenge is therefore to turn these beautiful qualitative pictures into quantitative ones. That has proved a very formidable task.
There are two parts in this challenge.
On the one hand, we need to be able to accurately read off the CFT and its observables from the tensor network.
On the other, we have to identify gravitational observables in the tensor network.
The former is hard if we build the tensor network using tensors engineered to recover nice bulk properties, such as the perfect/random tensor networks. In those constructions, it is unclear what the boundary CFT is.
The latter task is always hard, whether we start with a tensor network with known or unknown CFT connections.
One needs to identify matter excitations in the bulk, and also the metric of the background space-time.
Identification of matter content can be done in some toy models \cite{Bhattacharyya:2017aly, Hung:2019zsk} so that Witten diagrams emerge to some extent. But these results are far from satisfactory in most cases where the tensor network as a discretization breaks most of the symmetries of AdS space.
The hint for geometrical data lies mostly in the Ryu-Takayanagi formula that connects entanglement with areas of minimal surfaces.  But practically, to extract the metric from it remains cumbersome, particularly if the theory deviates from the fixed point, and that the corresponding tensor network describes a background that deviates from the pure AdS space. Some attempts based on the Fisher metric between density matrices and correlation functions of the CFT \cite{Mollabashi:2013lya, Lee:2015vla, Miyaji:2015fia, Caputa:2017yrh} have been made, and they resemble known gravitational solutions to different extent.

To make progress, one needs to put these proposals to further tests by showing that the dynamics of gravity should emerge from the tensor network description of the CFT as well.
All the difficulties described above mingle in this task. There are some progresses based on extremalising relative entropies \cite{Bhattacharya:2013bna, Blanco:2013joa, Faulkner:2013ica,Faulkner:2014jva, Jacobson:2015hqa} and complexities, but the results either require various assumptions (such as the applicability of the RT formula \cite{Faulkner:2014jva}, or certain behaviour of the modular hamiltonian of the matter fields \cite{Jacobson:2015hqa}), or that they are not explicitly covariant, which is a common issue that plague tensor network reconstruction of CFT states.See also some progress obtaining the Einstein equation using the bit-threads approach \cite{Freedman:2016zud, Agon:2020mvu}.

There is an ideal testing ground where quantitative computations can be more readily made. If making overly simplistic choice of tensors has little hope of recovering well-behaved CFT's and AdS bulks, then perhaps the second best option is to work with a simplified version of the AdS/CFT correspondence. The p-adic AdS/CFT \cite{Gubser:2016guj,Heydeman:2016ldy} is such a perfect arena where the correspondence preserves much of the essence of the AdS/CFT. \footnote{More recent progress see \cite{Ebert:2019sr,Jepsen:2019svc,Qu:2019tyi,Qu:2018ned,Parikh:2019dvm, Garcia-Compean:2019jvk}. }
The boundary CFT is sufficiently simple that can be precisely reconstructed by a tensor network covering the dual bulk -- which is naturally {\it discrete} in this case \cite{Bhattacharyya:2017aly,Hung:2019zsk, Heydeman:2018qty}.  We note that this construction recovers the CFT partition function rather than a state at a given time slice, and so it naturally avoids issues of covariance mentioned above.
Many of the well known items in the AdS/CFT dictionary, such as the correspondence between primary operators and bulk fields, their bulk-boundary propagators, the HKLL formula \cite{Hamilton:2006fh, Hamilton:2005ju} and Witten diagrams quantitatively emerge from the tensor network representation.
It remains to show that this tensor network can describe space-times deviated from the pure {\it Bruhat-Tits tree} geometry, the p-adic analogue of pure AdS space, and that the interplay between matter and geometry follows rules that can be interpreted as the Einstein equation on a discrete graph.

In this paper, which is an extended companion to \cite{paper1}, we will take the tensor network construction of the p-adic CFT partition function in \cite{Hung:2019zsk}, and achieve precisely this goal, perturbatively away from the pure Bruhat-Tits geometry.
Our procedure comes in 6 steps.
\begin{enumerate}
\item{Define the notion of bulk operator insertion in the tensor network so that it is consistent with the AdS/CFT dictionary, including the correct bulk-bulk propagator and the HKLL formula.  The bulk and boundary correlation functions where the boundary conditions correspond to the CFT fixed points are consistent with a quantum field theory living on the Bruhat-Tits tree. We can read off the bulk action as well.
This is reviewed and expanded in section 2.  }

\item{Obtain a deformed geometry by picking appropriate boundary conditions of the tensor network. This has been defined and partially studied in our previous sequel \cite{Hung:2019zsk}. These boundary conditions are the direct analogue of choosing non-trivial boundary conditions of bulk scalar fields which corresponds to turning on non-trivial relevant operators in the CFT to drive an RG flow.  We can then read off the expectation value of bulk matter fields from the tensor network when the boundary conditions of the tensor network have been deformed.
This is explained in section 2.}

\item{In the deformed geometry, one needs to define the notion of distance. Rather than making guesses, we assume that it has to depend on tensor network data locally and isotropically. In the case where deviations from the CFT fixed point boundary condition is small, this dependence can be expressed as a power series of the deformation parameters. We treat the expansion coefficients, with symmetries following from the locality and isotropy assumptions, as unknowns to be determined. This is discussed in section 3.  }

\item{The notion of graph curvature is generically a local function of edge distances on the graph \cite{Yau1,Ollivier,Gubser:2016htz}. We also stay agnostic about the precise definition of the graph curvature, and consider an expansion of the curvature as a power series of deviation of edge distances from the {\it pure Bruhat-Tits} space. The coefficients are again constrained by locality and isotropy, but otherwise kept as unknowns. This is discussed in section 3.  }

\item{Having a tentative assignment of curvature and expectation values of bulk operators, an emergent Einstein equation is a relation between these data. To systematically look for this relation, we construct an action that encodes the bulk data. We have the notion of Einstein Hilbert term given the assignment of curvature discussed above. Then, on the part of the matter fields, we covariantize the matter action previously read off from the tensor network when the bulk is a {
\it pure Bruhat-Tits} geometry. Again we will expand their dependence on the edge lengths as a power series. This is discussed in section 4.
 }

 \item{With the graph curvature and covariant matter action in place, we can obtain a graph Einstein equation by varying the edge lengths.
 We substitute the expectation values of the bulk scalar fields and also the edge lengths as a power series of boundary conditions into the equations of motion. Since the boundary conditions are chosen to be arbitrary, assuming that the Einstein equation is satisfied  becomes a very stringent constraint on the unknown parameters that we have introduced thus far. In fact this is an over-determined system that is not guaranteed to have any solutions at all.  Amazingly, a solution exists, and that the unknown coefficients can be determined uniquely (up to some overall normalizations). The resultant graph curvature as a function of edge lengths recovers the proposal in the mathematics literature! This is discussed in section 4. }
\end{enumerate}

A review of p-adic CFT has appeared in many places. Since the current paper is focussed on bulk physics, we will relegate a brief review to the appendix \ref{padiccft} for completeness and for setting notations for the tensor network.

We will conclude in section 5.

\section{Tensor network reconstruction of p-adic AdS/CFT -- RG fixed point and deformations}

\begin{figure}
	\centering
	\includegraphics[width=0.3\linewidth]{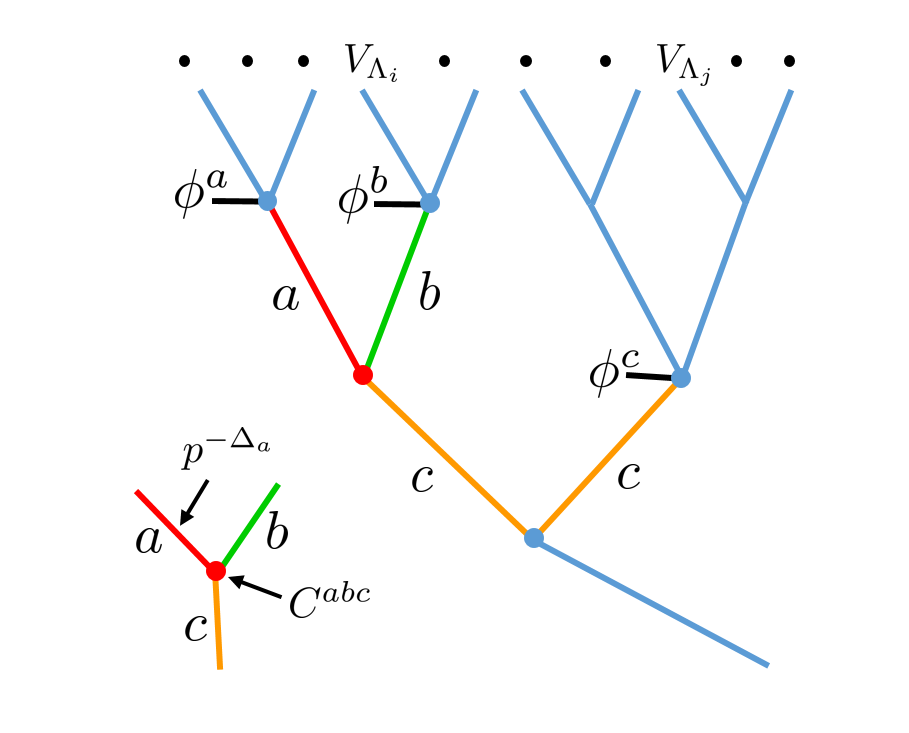}
	\caption{The tensor network representation of a $p=2$-adic CFT. The diagram depicts three bulk operator insertion. When these bulk insertions are pushed to the asymptotic boundary they are equivalent to boundary insertions. The boundary condition $V_{\Lambda_i}$ here are chosen to be the fixed point tensor $V^a_f = \delta^a_1$.
	Each vertex tensor is given by $C^{abc}$ and each edge of the tensor with index $a$ is weighted by $p^{-\Delta_a}$. }
	\label{fig:ggc}
\end{figure}

A tensor network that recovers the p-adic CFT partition function was introduced in \cite{Hung:2019zsk}. (For completeness a very brief review of p-adic CFTs is included in Appendix \ref{padiccft}.)
It is a tensor network that covers the Bruhat-Tits tree.
The Bruhat-Tits tree is an infinite tree graph whose isometry is given by the conformal group of the p-adic CFT. Each vertex has $p+1$ legs. The case for $p=2$ is depicted in Fig. \ref{fig:ggc}.

 For concreteness, we will focus on a 1-dimensional p-adic CFT that lives in the p-adic line $x\in Q_p$, where $Q_p$ is the p-adic number fields.  Although as emphasized before \cite{Gubser:2016guj}, generalization to $n$ dimensions corresponds to replacing $Q_p$ by a field extension $Q_{p^n}$. The bulk becomes a $p^n+1$ valent tree, and all expressions basically change by replacing $p \to p^n$. Therefore without loss of generality and to avoid clutter we will take $n=1$ in the rest of this paper.

 The tensor network constructed in \cite{Hung:2019zsk} puts a tensor at every vertex of the Bruhat-tits tree. The indices of these tensors are labeled by the primary operators of the CFTs. The value of the tensor $T_{a_1 \cdots a_{p+1}}$ is given by the fusion tree of $p+1$ primaries, with fusion coefficients given by the OPE coefficients of the CFT  (see appendix \ref{padiccft}).  i.e.
 \be
 T^{a_1 a_2 \cdots a_{p+1}} = \sum_{b_1\cdots b_{p-2}} C^{a_1 a_2 b_1} C^{b_1 a_3 b_2}  \cdots C^{b_{p-2} a_p a_{p+1}}.
 \ee
 In the special case where $p=2$, we simply have $T^{a_1 a_2 a_3} = C^{a_1 a_2 a_3}$.
 Tensors sitting at two vertices connected by an edge have one paired index contracted, with the sum over the index weighted by $p^{-\Delta_a}$, where $\Delta_a$ is the conformal dimension of the primary operator $\mathcal{O}_a$ in the p-adic CFT.

 \subsection{Boundary conditions and the bulk field theory}

 The asymptotic boundary of the Bruhat-Tits tree is the $Q_p$ line.
 The tensor network has to be cutoff near the asymptotic boundary of the Bruhat-Tits tree -- analogous to the cutoff introduced in AdS space.
 The dangling legs of the tensors are contracted with a reference vector
 \be |V_f\rangle \equiv   \sum_a V_f^a |a\rangle, \qquad V_f^a \equiv \delta^a_1.
 \label{eq:refV}
 \ee
  i.e. it projects the dangling legs to the identity operator label.  The tensor network evaluates to a number for such boundary conditions. This number is interpreted as the (normalized) partition function of the p-adic CFT.
To insert operator $\mathcal{O}_a(x)$ into the partition function, the dangling leg located at $x$ is projected to label $a$ instead.

One can see that the computation of CFT correlation functions naturally reduces to sums of {\it Witten diagrams}, constructed from bulk-boundary propagators $G_{a}(x_i, v)$ meeting at bulk vertices $v$. These bulk-boundary propagators are given by
\be
 G_a(x, v) = \frac{\zeta_p(2\Delta_a)}{p^{\Delta_a}} p^{-{\Delta_a}d(x,v)},
\ee
where $d(x,v)$ counts the number of edges connecting the boundary vertex linked to $x$ and the bulk vertex $v$.
Let us emphasize here that the result of the correlation functions of the CFT is read-off directly from the tensor network without reference to any action. And yet, it has the appearance of being composed of this propagators which are solutions to the graph Klein-Gordon equation
\begin{align} \label{eq:Greenfn}
&(\Box_v  + m_a^2)G_a(x,v) = \delta_{x,v}, \qquad    m_a^2 = -\frac{1}{\zeta_p(\Delta_a-1)\zeta_p(-\Delta_a)}, \qquad
  \zeta_p(s) \equiv \frac{1}{1-p^{-s}}. \\
&\Box_v \phi(v) \equiv \sum_{u\sim v} ( \phi(v)- \phi(u)).
\end{align}

The result of the tensor network thus implies the existence of a massive bulk field $\phi_a$ in 1-1 correspondence with each CFT primary $\mathcal{O}_a$.
We can define the notion of bulk operator insertion at a bulk vertex $v$ by fusing an extra leg with label $a$ to the bulk vertex. For fusion rules that are commutative such that $C^{abc}$ is completely symmetric in the three indices, we can simply stick an extra bulk leg to the vertex $v$ without worrying about the order of the fusion.  This is illustrated in Fig. \ref{fig:ggc}.
The computation of a three point correlation function of the CFT is also illustrated there. When these bulk insertions are pushed to the boundary they become boundary operator insertions. i.e. This definition of bulk operator insertion ensures that the {\it extrapolation dictionary} -- where boundary operator corresponds to moving a bulk operator towards the asymptotic boundary -- is automatically realized.

One can thus compute bulk correlation functions, which would be given by sums of {\it Witten diagrams} constructed from bulk-bulk propagators meeting at other bulk vertices. The bulk-bulk propagator is simply given by the same $G_a$ in (\ref{eq:Greenfn}), with the boundary vertex label $x$ moved to the bulk.

By construction, each bulk vertex where propagators meet gains a factor of the OPE coefficients. A three point vertex is weighted by $C^{abc}$.

These suggest that at least where the operator insertion at the boundary is sparse, the tensor network can be described by an emergent bulk quantum field theory with action given by
\begin{eqnarray}
  S_m &=& \sum_{\langle xy\rangle}\frac{1}{2}(\phi^a_x-\phi^a_y)^2+\sum_x \frac{1}{2}m_a^2(\phi^{a}_x)^2+\mathcal{O}(\phi^3),
\end{eqnarray}
where $x$ denotes the vertices, and $a$ denotes different fields whose mass is $m_a$. The $a$ labelling the a primary is summed over though not shown explicitly.
More accurately speaking, since the vertex where propagators meet in the tensor network is unique, it exactly matches with the semi-classical limit of the field theory, one where the masses $m_a$ approach infinity, or equivalently, $\Delta_a \to \infty$, so that the interaction vertex would be fixed at the intersection of geodesics.

To set the normalization rigorously, simple insertion of legs at a bulk vertex $x$ corresponds to inserting a field $\tilde \phi_x$ that is related to the canonically normalized $\phi_a(x)$ appearing in the above action by
\begin{eqnarray}
  \tilde{\phi}^a_x &\equiv& \left(\frac{\zeta_p(2\Delta_a)}{p^{\Delta_a}}\right)^{-\frac{1}{2}}\phi^a_x,
\end{eqnarray}
so that the two point function read off from the insertion of extra bulk legs is simply given by

\begin{eqnarray}
  \langle \tilde{\phi}^a_x \tilde{\phi}^a_y \rangle = p^{-\Delta_a d(x,y)}
\end{eqnarray}

In this normalization, the action is expressed as
\begin{eqnarray}
  S_m &=& \sum_{\langle xy\rangle}\frac{\zeta_p(2\Delta_a)}{2p^{\Delta_a}}(\tilde{\phi}^a_x-\tilde{\phi}^a_y)^2+\sum_x \frac{\zeta_p(2\Delta_a)}{2p^{\Delta_a}}m_a^2(\tilde{\phi}^{a}_x)^2+\mathcal{O}(\tilde{\phi}^3).
\end{eqnarray}
One can rewrite the summation over vertices to a sum over edges, which gives
\begin{eqnarray}
  S_m &=& \sum_{\langle xy\rangle}\frac{\zeta_p(2\Delta_a)}{2p^{\Delta_a}}(\tilde{\phi}^a_x-\tilde{\phi}^a_y)^2+\sum_{\langle xy\rangle} \frac{\zeta_p(2\Delta_a)}{2p^{\Delta_a}}\frac{1}{p+1}m_a^2((\tilde{\phi}^{a}_x)^2+(\tilde{\phi}^{a}_y)^2)+\mathcal{O}(\tilde{\phi}^3).  \label{eq:Sm1}
\end{eqnarray}
The factor $1/(p+1)$ is to cancel the $p+1$ times overcounting since each vertex attaches to $p+1$ edges.

 \subsection{RG flow via deformed boundary conditions}

 RG flow of the p-adic CFT was considered in \cite{Hung:2019zsk}.  In the tensor network, the RG flow of the CFT can be driven by changing the boundary conditions at the asymptotic boundary. Instead of projecting the boundary leg to the vector $V_{f}^a $ defined in (\ref{eq:refV}), we pick instead
 \be
 |V_{\Lambda}\rangle \equiv \sum_a V_\Lambda^a |a\rangle
 \ee
which generically turns on other primaries at the boundary.  The subscript $\Lambda$ labels the cutoff surface on the tensor network at which these reference vectors are inserted. The RG flow considered in \cite{Hung:2019zsk} respects translation invariance along $Q_p$. In that case, all boundary legs are projected to the same $|V_\Lambda\rangle$. More generally, one can pick different $|V_{\Lambda_i}\rangle$
for every boundary leg $x_i$.

We note that $p$ such boundary vectors would be fed to the same vertex tensor at the cutoff surface, which would then return a new vector $|V_{\Lambda-1}\rangle$:
\be
|V_{\Lambda-1} \rangle \equiv \sum_{a_{p+1}}V^{a_{p+1}}_{\Lambda-1} |a_{p+1}\rangle = \sum_{a_1\cdots a_p, a_{p+1}} p^{-(\Delta_{a_1}+ \cdots \Delta_{a_p})} T^{a_1\cdots a_p a_{p+1}} V_{\Lambda_1}^{a_1} \cdots V_{\Lambda_p}^{a_p} | a_{p+1}\rangle
\ee
i.e. $\Lambda-1 $ labels the legs one step away from the cutoff surface.

Note that the vector $|V_f\rangle$ is in fact a fixed point under this flow, explaining how a CFT partition function is recovered as fixed point vectors in the tensor network. For other choices of boundary vectors $|V_\Lambda\rangle$, a flow is driven down the network, so that different parts of the network carry different weights. Homogenous boundary conditions have been studied quite generally in \cite{Hung:2019zsk}, where we show that the flow would eventually lead to a new fixed point vector, corresponding to a CFT driven by relevant perturbations that eventually reaches another new CFT. The geometry in the interior of the network would thus resemble {\it pure Bruhat-Tits geometry} again, where the vertices eventually contribute equally to the partition function.
That the actual weights of the tensors contributing differently is suggestive of a varying metric in the tree. How the metric should depend on the tensors is the crux of reading off geometry and subsequently gravity from the tensor network. We will pick up this problem in the next section.

This is a highly non-linear flow and it is difficult to keep analytic control. To make further progress, we will consider small deviations from the fixed point vector $|V_f\rangle$.
i.e. In the following, we will consider
\be \label{eq:dV}
V^a_{\Lambda_i} = \delta^a_1 + \lambda v_{\Lambda_i}^a,
\ee
and treat $\lambda$ as a small parameter in a power series expansion around the CFT fixed point. Note that the subscript ``$\Lambda_i$'' denotes the position $i$ at the cutoff surface. At first sight, it may appear that a translation invariant boundary condition is simpler. But as we are going to see below, by assuming the most general perturbations  $\lambda_{\Lambda_i}$ at the boundary, it gives the strongest constraint of the emergent Einstein equation.

\subsection{Local data of the tensor network and expectation values }

\begin{figure}
	\centering
	\includegraphics[width=0.5\linewidth]{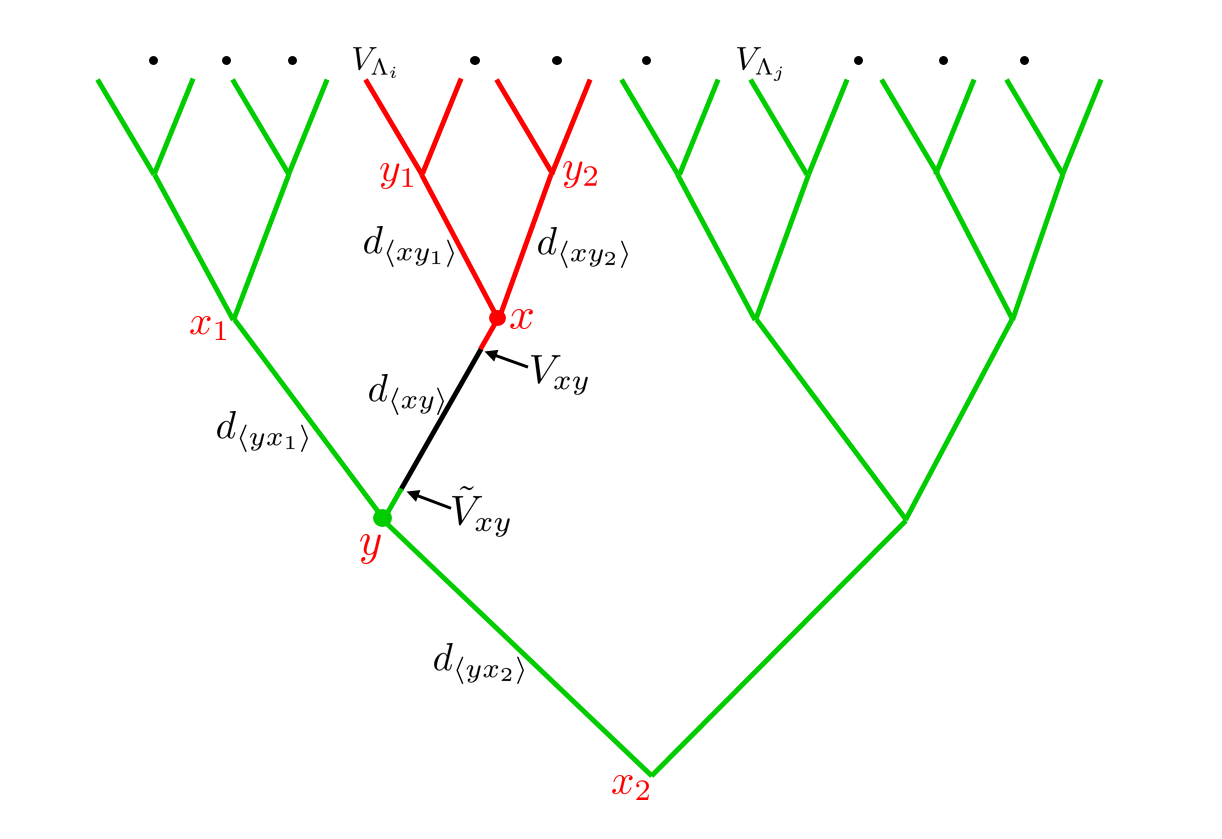}
	\caption{Vector $V^a_{xy}$ and $\tilde V^a_{xy}$ following from the contraction of tensors {\it above}  (colored red) and  {\it below} (colored green) the edge $\langle xy\rangle$ respectively.  The boundary conditions  $V_{\Lambda_i}$ generically deviate from the fixed point vector. The curvature of the patch centered at $x$ depends on the graph distances $d_{\langle x y_i\rangle} $ symmetrically. }
	\label{fig:redgreen}
\end{figure}

\begin{figure}[htbp!]
\centering
\subfigure[]{
\label{tensor}
\includegraphics[width=0.45\textwidth]{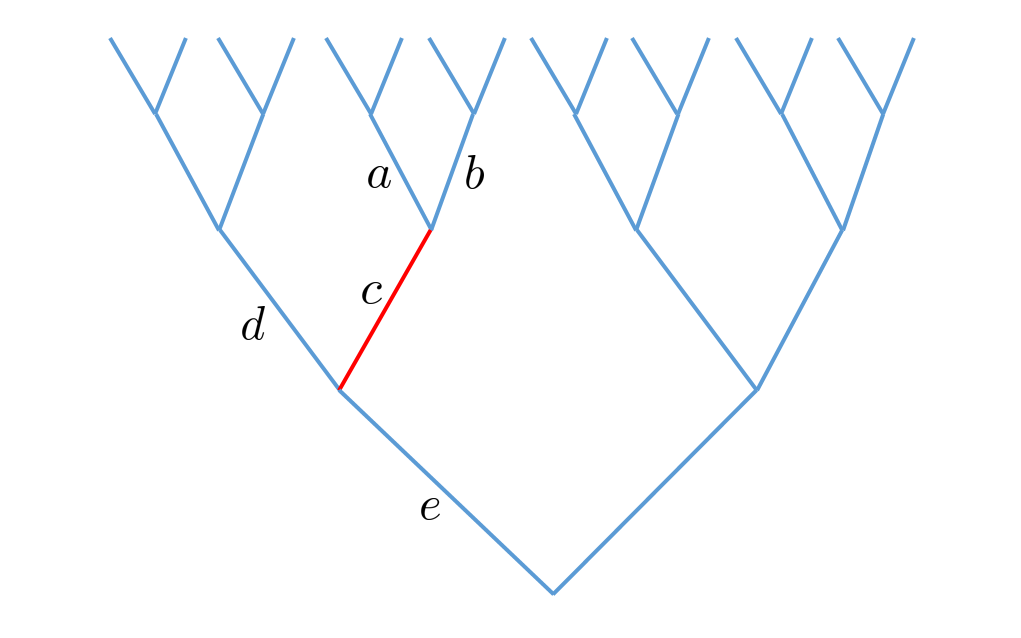}}
\subfigure[]{
\label{edge}
\includegraphics[width=0.3\textwidth]{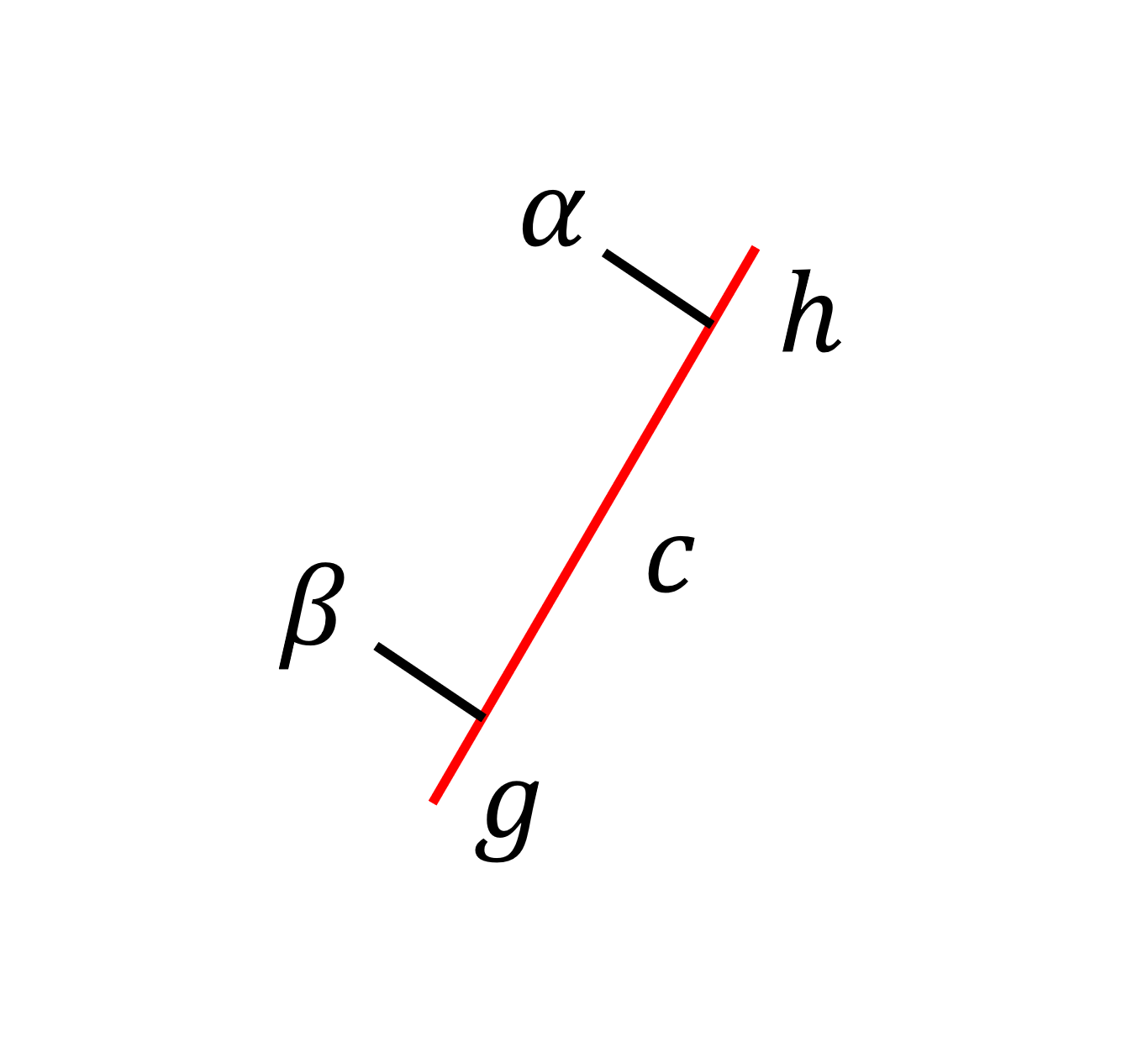}}
\caption{(a):Here we demonstrate the $p=2$ case for simplicity.
(b):The two black legs are bulk legs. We insert them near the vertices to make the labelling more transparent. They should be understood as inserted on the vertices}

\end{figure}

Consider the $p$-adic tensor network as shown in Fig. \ref{tensor}. Expanded in the contributing tensors, it takes the form
 \begin{eqnarray}
  Z= \sum_{\dots,a,b,c,d,e,\dots}\dots p^{-\Delta_a}p^{-\Delta_b}C^{abc}p^{-\Delta_c}C^{cde}p^{-\Delta_d}p^{-\Delta_e}\dots.
 \end{eqnarray}
We can rewrite $Z$ as
\begin{eqnarray}
  Z &=& \sum_c V^c p^{-\Delta_c}\tilde{V}^c,
\end{eqnarray}
defining
\begin{eqnarray}
  V^c &\equiv& \sum_{\dots,a,b}\dots p^{-\Delta_a}p^{-\Delta_b}C^{abc},\\
\tilde{V}^c &\equiv&\sum_{d,e,\dots}C^{cde}p^{-\Delta_d}p^{-\Delta_e}\dots.
\end{eqnarray}
When one is restricted on the red edge, $V^a$ and $\tilde{V}^a$ carry all the information in the tree.
They follow from contracting all the tensors {\it above} or {\it below} the edge. This is illustrated in Fig. \ref{fig:redgreen}.

Using these definitions, the two point function on this edge regardless of boundary condition can always be written in terms of the pair of $V^a$ and $\tilde{V}^a$
\begin{eqnarray}
  \langle \tilde \phi_\alpha \tilde \phi_\beta \rangle &=& \sum_{h,c,g} V^h C^{hc\alpha}p^{-\Delta_c}C^{cg\beta}\tilde{V}^g,
\end{eqnarray}
as shown in Fig. \ref{edge}.

In the $p$-adic vacuum where the boundary legs are projected to the fixed point vector $|V_f\rangle$, as expected, $V^a$ and $\tilde{V}^a$ are
both equal to the fixed point vector.
\begin{eqnarray}
  V^a = \delta_1^a,\; \tilde{V}^a=\delta_1^a,
\end{eqnarray}

Now, anticipating what will be needed when we begin considering curvatures, let us in particular consider the following local patch of the $p$-adic tensor network as shown in Fig. \ref{part}. The edge $xy$ is connected with $\{xy_i\}$ and $\{x_iy\}$. When one is restricted within this part, we can express all relevant information describing this patch from the collection $\{V_i^a\}$ and $\{\tilde{V}_i^a\}$ located at each of the out-stretching legs. One can see  that for edge $xy_i$, its $V^a$ is $V_i^a$, while for edge $x_iy$, its $\tilde{V}^a$ is $\tilde{V}_i^a$. Perturbed around the vacuum under generic boundary conditions (\ref{eq:dV}), we have
\begin{eqnarray}
  V_i^a &=&\delta_1^a+ \omega_i^a, \qquad \omega_i^a \equiv \lambda_i^a+\eta_i^a+\mathcal{O}(\lambda^3),   \nonumber \\
\tilde{V}_i^a &=&\delta_1^a+\tilde{\omega}_i^a  , \qquad \tilde{\omega}_i^a \equiv \tilde{\lambda}_i^a+\tilde{\eta}_i^a+\mathcal{O}(\lambda^3), \label{eq:Cexpand}
\end{eqnarray}
where $\lambda_i^a,\tilde{\lambda}_i^a\sim \mathcal{O}(\lambda)$, $\eta_i^a,\tilde{\eta}_i^a\sim\mathcal{O}(\lambda^2)$, and $\lambda\ll1$.

\begin{figure}[htbp!]
\centering

\label{part}
\includegraphics[width=0.6\textwidth]{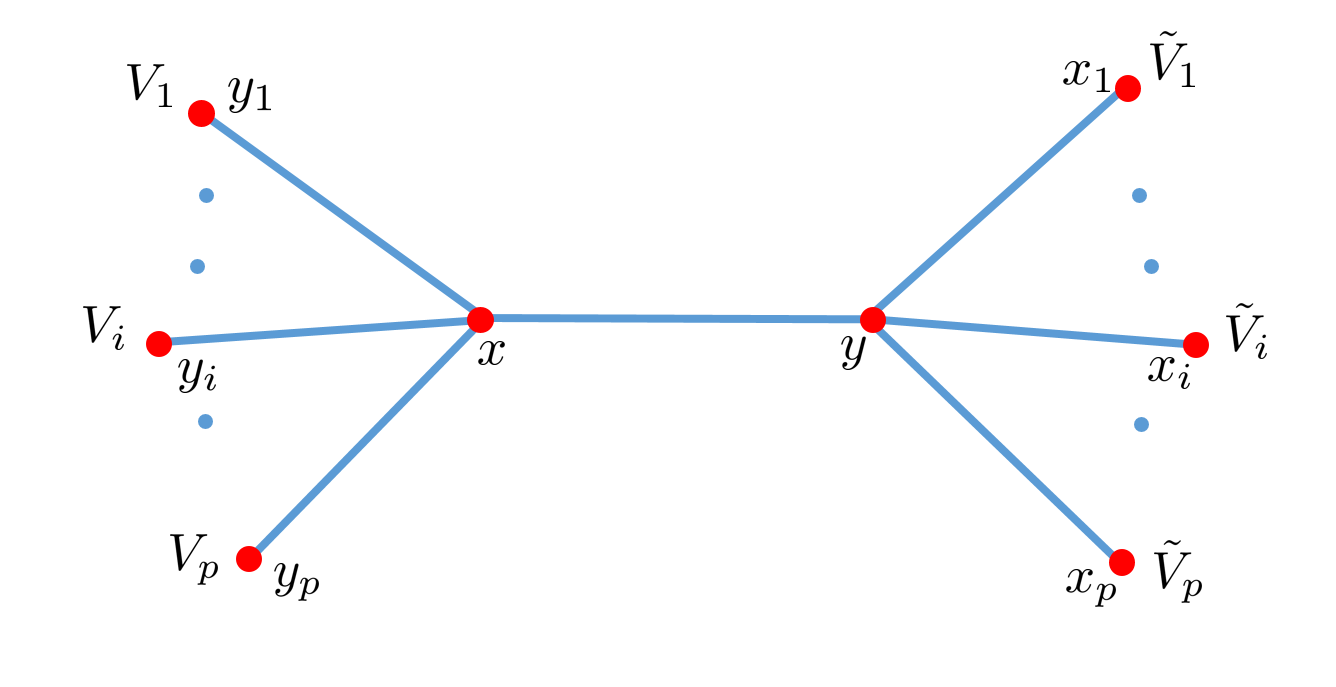}

\caption{The edge $xy$ is connected with $\{xy_i\}$ and $\{x_iy\}$. When one is restricted within this part, $\{V_i^a\}$ and $\{\tilde{V}_i^a\}$ encode all the information needed.}
\label{part}

\end{figure}

Reading off from the tensor network, we have
\begin{eqnarray} \label{eq:lambda_def}
\label{phixa}
 \langle \tilde{\phi}_x^a\rangle &=&\lambda^a p^{-\Delta_a}+\tilde{\lambda}^a p^{-2\Delta_a}+\mathcal{O}(\lambda^2), \\
  \label{phiya}
 \langle  \tilde{\phi}_y^a \rangle &=& \tilde{\lambda}^a p^{-\Delta_a}+\lambda^a p^{-2\Delta_a}+\mathcal{O}(\lambda^2),
\end{eqnarray}
where $\lambda^a,\tilde{\lambda}^a$ are defined by
\begin{eqnarray}
  \lambda^a &\equiv& \sum_{i=1}^p \lambda^a_i,\\
  \tilde{\lambda}^a &\equiv& \sum_{i=1}^p \tilde{\lambda}^a_i.
\end{eqnarray}

\section{Distances and graph curvature from the tensor network}

We have discussed the local data that is encoded in the tensor network in detail in the previous section.
In this section, we would like to explain how one should assign distances to edges and graph curvatures to patches based on the local data obtained from the tensor network.

\subsection{Edge distances from the tensor network}

As discussed in the previous section using Fig. \ref{edge}, all the information carried by an edge $e$ is completely captured by $V_{e}^a$ and $\tilde{V}_{e}^a$ at the two ends of the edge.
So it's natural to propose that the edge length $d_e$ is also determined by $V_{e}^a$ and $\tilde{V}_{e}^a$.
It is a bit challenging to obtain the complete dependence of the edge length on $V_{e}^a$ and $\tilde{V}_{e}^a$.
However, if we are working with the perturbative limit, where $V_{e}^a$ and $\tilde{V}_{e}^a$ admit a perturbation expansion of the form (\ref{eq:Cexpand}):
\begin{eqnarray}
  V_{e}^a &=&\delta^a_1+w^a_e= \delta^a_1+ \lambda^a_e+\eta^a_e+\mathcal{O}(\lambda^3),\\
  \tilde{V}_{e}^a &=&\delta^a_1+\tilde{w}^a_e= \delta^a_1+ \tilde{\lambda}^a_e+\tilde{\eta}^a_e+\mathcal{O}(\lambda^3).
\end{eqnarray}
then the edge distance $d_e$ also admits an expansion in $\lambda^a_e, \tilde{\lambda}^a_e, \eta^a_e, \tilde{\eta}^a_e$.

But first, we note that when the boundary condition corresponds to the fixed point vector, every vertex and thus every edge gives
equal contribution to the partition function. Therefore precisely at the fixed point, all edge distances are equal, and we can conveniently set it to 1.
Therefore, under perturbation away from the fixed point, the edge length at each edge is also admitting small deviation from unity as follows:

\begin{eqnarray}
  d_{e} &=& 1+j_{e}
\end{eqnarray}
with $j_{e}\ll 1$, and this deviation should depend on the local data on the tensor network.
Therefore we can also expand it in powers of $\lambda$ as follows:
\begin{eqnarray}
  j_e &=&A^a(\omega_e^a+\tilde{\omega}_e^a)+B^{ab}(\omega_e^a\omega_e^b+\tilde{\omega}_e^a\tilde{\omega}_e^b)+C^{ab}\omega_e^a\tilde{\omega}_e^b+\mathcal{O}(\omega^3)  \nonumber\\
  &=& A^a(\lambda_e^a+\tilde{\lambda}_e^a)+B^{ab}(\lambda_e^a\lambda_e^b+\tilde{\lambda}_e^a\tilde{\lambda}_e^b)+C^{ab}\lambda_e^a\tilde{\lambda}_e^b+A^a(\eta_e^a+\tilde{\eta}_e^a)+\mathcal{O}(\lambda^3),\;\;\;\;\;
 \label{de}
\end{eqnarray}
with $B^{ab},C^{ab}$ symmetric matrices that parametrise this expansion. Here we have used the symmetry of $\omega_e^a\leftrightarrow\tilde{\omega}_e^a$, i.e. $V_e^a\leftrightarrow\tilde{V}_e^a$, since the edge should depend symmetrically on the information at its two ends. These unknown parameters, as we are going to find out, are almost uniquely fixed if a consistent Einstein equation is to exist at all.

\subsection{Graph curvature from edge distances and the Einstein Hilbert action}
We would like to define graph Ricci curvatures given the set of edge distances we have now defined on the tensor network.
Graph curvatures have been considered in the mathematics literature \cite{Yau1,Ollivier,Gubser:2016htz, Huang:2020qik}.
Inspired by them, we consider the Ricci curvature $R(d_{xy_1},d_{xy_2},\dots,d_{xy_{p+1}})$ of a patch around a vertex $x$ to be a symmetric function of the lengths $d_{xy_i}$ of edges emanating from the vertex $x$.
For the precise dependence however, we first take the agnostic approach, and consider the most general expansion of the curvature on the edge distances perturbed away from unity. We expect that there is a regular expansion since the curvature should evaluate to a constant when all edges have equal lengths.

We parametrize the expansion of the graph Ricci curvature $R(d_{xy_1},d_{xy_2},\dots,d_{xy_{p+1}})$ as a power series in $\{j_{xy_i}\}$ as:
\begin{eqnarray} \label{eq:Ricci}
  R(d_{xy_1},d_{xy_2},\dots,d_{xy_{p+1}}) &=& a_0+a_1\sum_i j_{xy_i}+b\sum_i j^2_{xy_i}+c\sum_{i\neq k}j_{xy_i}j_{xy_k}+\mathcal{O}(j^3).
\end{eqnarray}
Here we have imposed symmetric dependence of all edges connected to $x$. These coefficients $a_0, b,c$
are dependence of curvatures on distances and they should be universal independent of the tensor network.

Now we would like to write down the analogue of the Einstein-Hilbert action on the graph.
On Riemann manifold the Einstein-Hilbert action takes the form:
\begin{eqnarray}
  S_{EH}[g_{\mu\nu}] &=& \int d^dx \sqrt{-g}(R+\Lambda),
\end{eqnarray}
with $\Lambda$ the cosmological constant and $R$ the Ricci scalar which is a function of the metric $g_{\mu\nu}$.

Having defined the graph scalar curvature in (\ref{eq:Ricci}), the Einstein-Hilbert action on the BT tree should take the general form:
\begin{eqnarray}
  S_{EH} &=& \sum_x R(d_{xy_1},d_{xy_2},\dots,d_{xy_{p+1}})+\sum_{\langle xy\rangle} d_{xy} \Lambda, \label{eq:EH}
\end{eqnarray}
where the first sum runs over all bulk vertices $x$,  and $\sum_{\langle xy\rangle}$ indicates a sum over edges.
The second term is proposed to mimic the cosmological constant term. In the continuous case, the volume form is given by $d^dx \sqrt{-g}$.
Therefore it is natural to propose that
\begin{eqnarray}
 d^dx \sqrt{-g} &\sim& d_{xy},  \label{eq:vol}
\end{eqnarray}
which leads to the second term in (\ref{eq:EH}).

\section{Covariant matter action and a graph Einstein equation}

Having obtained an ansatz of edge distances and graph curvature, we would like to consider covariantizing the matter action, and eventually obtain a graph Einstein equation.
\subsection{A covariant matter action}
The matter action read off from the tensor network  when the boundary conditions are taken to be the fixed point vectors except at isolated locations of operator insertion is given by (\ref{eq:Sm1}). This is essentially the action for pure BT space.

We would like to {\it covariantize} this action, to couple edge lengths to the matter fields. This has been considered in \cite{Gubser:2016htz}. However we find the ansatz quite restrictive there, and we would like to write down a more general ansatz that mimics closely the covariant coupling of matter to the background in the continuous case, although very much like in our treatment of the edge lengths and graph curvature, the ansatz allows for a collection of parameters that cannot be fixed purely by symmetry and locality considerations.

Recall (\ref{eq:vol}), then
\begin{eqnarray}
 \int d^dx \sqrt{-g}f(x) &\rightarrow& \sum_{\langle xy\rangle}d_{xy}f_{xy}.
\end{eqnarray}
A natural ansatz for the covariantized matter action $S^{cov}_m$ therefore takes the following form:
\begin{eqnarray}
  S^{cov}_m &=& \sum_{\langle xy\rangle}d^k_{xy}\frac{\zeta_p(2\Delta_a)}{2p^{\Delta_a}}(\tilde{\phi}^a_x-\tilde{\phi}^a_y)^2+\sum_{\langle xy\rangle}d_{xy} \frac{\zeta_p(2\Delta_a)}{2p^{\Delta_a}}\frac{1}{p+1}m_a^2((\tilde{\phi}^{a}_x)^2+(\tilde{\phi}^{a}_y)^2)+\mathcal{O}(\tilde{\phi}^3),
\end{eqnarray}
where $k$ is some constant we can't determine now since $(\tilde{\phi}^a_x-\tilde{\phi}^a_y)^2$ may contribute additional $1/d^2_{xy}$.
Writing down the general $\mathcal{O}(\tilde{\phi}^3)$ term explicitly, $S^{cov}_m$ becomes
\begin{eqnarray}
\nonumber
 S^{cov}_m&=&\sum_{\langle xy\rangle}d^k_{xy}\frac{\zeta_p(2\Delta_a)}{2p^{\Delta_a}}(\tilde{\phi}^a_x-\tilde{\phi}^a_y)^2+\sum_{\langle xy\rangle}d_{xy} \frac{\zeta_p(2\Delta_a)}{2p^{\Delta_a}}\frac{1}{p+1}m_a^2((\tilde{\phi}^{a}_x)^2+(\tilde{\phi}^{a}_y)^2)\\
 &\;&+\sum_{\langle xy\rangle}\left(h(d_{xy})H^{abc}(\tilde{\phi}^a_x\tilde{\phi}^b_x\tilde{\phi}^c_x+\tilde{\phi}^a_y\tilde{\phi}^b_y\tilde{\phi}^c_y)+
 r(d_{xy})R^{abc}(\tilde{\phi}^a_x\tilde{\phi}^b_x\tilde{\phi}^c_y+\tilde{\phi}^a_y\tilde{\phi}^b_y\tilde{\phi}^c_x)\right)+\mathcal{O}(\tilde{\phi}^4),\;\;\;\;
\label{actionm}
\end{eqnarray}
where $h(d_{xy}),r(d_{xy})$ are functions of $d_{xy}$, and $H^{abc}$ is totally symmetric, while $R^{abc}$ is symmetric in $a,b$.

Similar to our previous treatment of the graph curvature,  we can expand $h(d_{xy}),r(d_{xy})$ in powers of perturbations $j_{xy}$ of the edge lengths :
\begin{eqnarray}
  h(d_{xy}) &=& h_0+h_1 j_{xy}+h_2 j^2_{xy}+\dots, \\
  r(d_{xy}) &=& r_0+r_1 j_{xy}+r_2 j^2_{xy}+\dots.
\end{eqnarray}

\subsection{A graph Einstein equation}

We have proposed the ansatz for an Einstein-Hilbert action and a covariant matter action.
We are now ready to obtain a graph Einstein equation by varying the total action $S_{EH} + S^{cov}_m$ with respect to the edge lengths.
This gives
\begin{eqnarray}
  \frac{\delta S_{tot}}{\delta d_{xy}}=\frac{\delta S_{EH}}{\delta d_{xy}}+\frac{\delta S^{cov}_m}{\delta d_{xy}}=0, \;\;\textrm{i.e.}\;\; G+T=0,
\end{eqnarray}
with
\begin{eqnarray}
G\equiv\frac{\delta S_{EH}}{\delta d_{xy}} &=&  \Lambda+2a_1+4b j_{xy}+c(\sum_{\substack{i\\(y_i\neq y)}} j_{xy_i}+\sum_{\substack{i\\(x_i\neq x)}}j_{x_iy})+\mathcal{O}(j^2),
\label{G}\\
  T \equiv \frac{\delta S^{cov}_m}{\delta d_{xy}}&=&
  k\frac{\zeta_p(2\Delta_a)}{2p^{\Delta_a}}(\tilde{\phi}^a_x-\tilde{\phi}^a_y)^2+ \frac{\zeta_p(2\Delta_a)}{2p^{\Delta_a}}\frac{1}{p+1}m_a^2((\tilde{\phi}^{a}_x)^2+(\tilde{\phi}^{a}_y)^2)
  \nonumber\\
 &\;&+\left(h_1H^{abc}(\tilde{\phi}^a_x\tilde{\phi}^b_x\tilde{\phi}^c_x+\tilde{\phi}^a_y\tilde{\phi}^b_y\tilde{\phi}^c_y)+
 r_1R^{abc}(\tilde{\phi}^a_x\tilde{\phi}^b_x\tilde{\phi}^c_y+\tilde{\phi}^a_y\tilde{\phi}^b_y\tilde{\phi}^c_x)\right)+\mathcal{O}(\tilde{\phi}^4),
 \label{T}
\end{eqnarray}

One very important fact to note is that by considering the equations following from variation of $d_{xy}$,  the edge data involved in the equation span a patch that is precisely given by Fig. \ref{part}, which was anticipated there.

\subsection{Solving the Einstein constraints to order $\lambda^2$}

We have obtained from the tensor network expectation values of some scalar field $\tilde \phi_a$. We have also read off edge lengths and edge curvatures based on data of the tensor network. The issue at hand is that -- we have a solution in search of an equation relating these data.
Using very few assumptions, based on locality and also correlation functions of these scalar fields at the CFT fixed point, we have written down an ansatz of an action up to some undetermined coefficients, which led to an Einstein equation. Here we would like to ask the following question:
if the expectation values and geometrical data that we have read off from the tensor network are indeed related by the purported graph Einstein equation, what are the allowed values of the undetermined parameters so that the Einstein equation can indeed be satisfied, if at all?

We are therefore going to treat the Einstein equation as a constraint on the undetermined parameters.
The surprising result is that this over-determined system does have a unique solution (up to some overall normalizations) that naturally recovers results in the mathematics literature and satisfies all other consistency conditions.

We will demonstrate this order by order in the $\lambda$ expansion.

Plugging (\ref{phixa}), (\ref{phiya}) into (\ref{T}), we find that
\begin{eqnarray}
  T &=& \frac{p^{-3 \Delta_a} \left(k (p+1) \left(p^{\Delta_a}-1\right)^2+m_a^2 \left(p^{2 \Delta_a}+1\right)\right)}{2(p+1) \left(p^{\Delta_a}-1\right) \left(p^{\Delta_a}+1\right)} (\lambda^a\lambda^a+\tilde{\lambda}^a\tilde{\lambda}^a)\\
  &\;&+
  \frac{p^{-3 \Delta_a} \left(2 m_a^2 p^{\Delta_a}-k (p+1) \left(p^{\Delta_a}-1\right)^2\right)}{(p+1) \left(p^{\Delta_a}-1\right) \left(p^{\Delta_a}+1\right)}\lambda^a\tilde{\lambda}^a+\mathcal{O}(\lambda^3).
\end{eqnarray}

For the edge $xy_i$, its $V^a$ and $\tilde{V}^a$ can also be read off from the tensor network:
\begin{eqnarray}
  V^a &=& V^a_i=\delta_1^a+\lambda_i^a+\mathcal{O}(\lambda^2),\\
  \tilde{V}^a &=&\delta_1^a+(\lambda^a-\lambda^a_i)p^{-\Delta_a}+\tilde{\lambda}^ap^{-2\Delta_a}+\mathcal{O}(\lambda^2).
\end{eqnarray}
Similarly, for the edge $x_iy$, its $V^a$ and $\tilde{V}^a$ are given by
\begin{eqnarray}
  V^a &=& \delta_1^a+(\tilde{\lambda}^a-\tilde{\lambda}^a_i)p^{-\Delta_a}+\lambda^ap^{-2\Delta_a}+\mathcal{O}(\lambda^2), \\
  \tilde{V}^a &=&\tilde{V}^a_i=\delta_1^a+\tilde{\lambda}_i^a+\mathcal{O}(\lambda^2),
\end{eqnarray}
where we recall that the definition of $\lambda^a$ and $\tilde\lambda^a$ first appeared in (\ref{eq:lambda_def}).

For the edge $xy$, its $V^a$ and $\tilde{V}^a$ are given by
\begin{eqnarray}
  V^a &=& \delta_1^a+\lambda^a p^{-\Delta_a}+\mathcal{O}(\lambda^2),\\
\tilde{V}^a&=&\delta_1^a+\tilde{\lambda}^ap^{-\Delta_a}+\mathcal{O}(\lambda^2).
\end{eqnarray}

Plugging them into (\ref{de}) and recalling that $d_{xy}=1+j_{xy}$, we can obtain $j_{xy},\{j_{xy_i}\},\{j_{x_iy}\}$. Plugging all the $j$ into (\ref{G}), we find that
\begin{eqnarray}
  G = \Lambda+2a_1+A^a (\lambda^a+\tilde{\lambda}^a) p^{-2 \Delta_a} \left(4 b p^{\Delta_a}+c \left(p^{2 \Delta_a}+(p-1) p^{\Delta_a}+p\right)\right)+\mathcal{O}(\lambda^2).
\end{eqnarray}
To satisfy $G+T=0$ order by order in $\lambda$, we should have
\begin{eqnarray}
  \Lambda+2a_1 &=& 0,\\
  A^a&=&0,
\end{eqnarray}
since $T\sim \mathcal{O}(\lambda^2)$ and $\left(4 b p^{\Delta_a}+c \left(p^{2 \Delta_a}+(p-1) p^{\Delta_a}+p\right)\right)$ is zero for at most two $\Delta_a$. \footnote{The two solutions are $\Delta_a$ and $d-\Delta_a$, where $d$ is the dimension of the CFT which is conveniently taken to be one here. }

Then (\ref{de}) becomes
\begin{eqnarray}
  \label{de1}
  d_e = 1+B^{ab}(\lambda_e^a\lambda_e^b+\tilde{\lambda}_e^a\tilde{\lambda}_e^b)+C^{ab}\lambda_e^a\tilde{\lambda}_e^b+\mathcal{O}(\lambda^3),
\end{eqnarray}
which means we can work out the $\mathcal{O}(\lambda^2)$ term only with $\lambda_e^a, \tilde{\lambda}_e^a$. Plugging them into (\ref{G}), this time we find that
\begin{eqnarray}
\nonumber
  G &=&
  p^{-2 (\Delta_a+\Delta_b)} \left(4 b B^{ab} p^{\Delta_a+\Delta_b}+B^{ab} c \left(-2 p^{\Delta_a+\Delta_b}+p^{\Delta_a+\Delta_b+1}+p\right)+c C^{ab} p^{2 \Delta_a+\Delta_b}\right)(\lambda^a\lambda^b+\tilde{\lambda}^a\tilde{\lambda}^b)\\
  \nonumber
  &\;&+p^{-2 (\Delta_a+\Delta_b)} \left(4 b C^{ab} p^{\Delta_a+\Delta_b}+2 B^{ab} c (p-1) \left(p^{\Delta_b}+p^{\Delta_a}\right)+c C^{ab} \left(p^{2 \Delta_a}+p^{2 \Delta_b}\right)\right)\lambda^a\tilde{\lambda}^b\\
  &\;&+\sum_i \frac{c}{2} \left(2B^{ab}(1+p^{-\Delta_a-\Delta_b})-C^{ab} (p^{-\Delta_a}+p^{-\Delta_b})\right)(\lambda^a_i\lambda^b_i+\tilde{\lambda}^a_i\tilde{\lambda}^b_i)+\mathcal{O}(\lambda^3).
\end{eqnarray}

To satisfy $G+T=0$, firstly we should have
\begin{eqnarray}
  \sum_i \frac{c}{2} \left(2B^{ab}(1+p^{-\Delta_a-\Delta_b})-C^{ab} (p^{-\Delta_a}+p^{-\Delta_b})\right)(\lambda^a_i\lambda^b_i+\tilde{\lambda}^a_i\tilde{\lambda}^b_i) &=& 0,
\end{eqnarray}
since other terms in $G+T$ only depend on $\lambda^a,\tilde{\lambda}^a$. It is obvious that $c=0$ is one solution for it. That is too stringent. We are interested in $c\neq 0$ case which is more non-trivial, and this gives a constraint on the parameters
\begin{eqnarray}
  2B^{ab}(1+p^{-\Delta_a-\Delta_b})-C^{ab} (p^{-\Delta_a}+p^{-\Delta_b})=0,
\end{eqnarray}
which is already made symmetric in $a,b$ since it was contracted with a symmetric tensor $\lambda^a_i\lambda^b_i$. So we have
\begin{eqnarray}
  C^{ab} &=& \frac{2B^{ab}(1+p^{-\Delta_a-\Delta_b})}{p^{-\Delta_a}+p^{-\Delta_b}}.
\end{eqnarray}

Now comparing $G$ with $T$, we notice that $B^{ab}$ should be diagonal. And this gives
\begin{eqnarray}
\nonumber
  G &=& B^{aa} p^{-4 \Delta_a} \left(4 b p^{2 \Delta_a}+\frac{1}{2} c \left(-2 p^{3 \Delta_a}+2 p^{5 \Delta_a}+2 p^{\Delta_a+1}+2 p^{3 \Delta_a+1}\right) p^{-\Delta_a}\right)(\lambda^a\lambda^a+\tilde{\lambda}^a\tilde{\lambda}^a)\\
  &\;&+2 B^{aa} p^{-3 \Delta_a} \left(2 b \left(p^{2 \Delta_a}+1\right)+c \left(p^{2 \Delta_a}+2 p-1\right)\right)\lambda^a\tilde{\lambda}^a+\mathcal{O}(\lambda^3).
\end{eqnarray}
Then $G+T=0$ gives
\begin{eqnarray}
\nonumber
   &\;&B^{aa} p^{-4 \Delta_a} \left(4 b p^{2 \Delta_a}+\frac{1}{2} c \left(-2 p^{3 \Delta_a}+2 p^{5 \Delta_a}+2 p^{\Delta_a+1}+2 p^{3 \Delta_a+1}\right) p^{-\Delta_a}\right)\\ &=&-\frac{p^{-3 \Delta_a} \left(k (p+1) \left(p^{\Delta_a}-1\right)^2+m_a^2 \left(p^{2 \Delta_a}+1\right)\right)}{2(p+1) \left(p^{\Delta_a}-1\right) \left(p^{\Delta_a}+1\right)},\\
   \nonumber
\end{eqnarray}
and also
\begin{eqnarray}
   &\;&2 B^{aa} p^{-3 \Delta_a} \left(2 b \left(p^{2 \Delta_a}+1\right)+c \left(p^{2 \Delta_a}+2 p-1\right)\right)\\
   &=&-\frac{p^{-3 \Delta_a} \left(2 m_a^2 p^{\Delta_a}-k (p+1) \left(p^{\Delta_a}-1\right)^2\right)}{(p+1) \left(p^{\Delta_a}-1\right) \left(p^{\Delta_a}+1\right)} .
\end{eqnarray}
And $m_a^2$ can be solved by $b,c,k$:
\begin{eqnarray}
  m_a^2 &=& -k (p+1)-\frac{c k (p+1) \left(p^{1-\Delta_a}+p^{ \Delta_a}\right)}{2 b-c}.
  \label{m1}
\end{eqnarray}

Recall that for scalar field $\phi^a$ on the BT tree, its mass is related to the conformal dimension by
\begin{eqnarray}
\label{m2}
  m_a^2 &=& -\frac{1}{\zeta_p(\Delta_a-1)\zeta_p(-\Delta_a)}=-p-1+p^{1-\Delta_a}+p^{\Delta_a},
\end{eqnarray}
which is obtained by analyzing its Green's function.
 Comparing (\ref{m1}) with (\ref{m2}), we find that
\begin{eqnarray}
  k &=& 1,\\
  \frac{2b}{c}&=&-p,
\end{eqnarray}
which ensures that $k,b,c$ do not depend on $\Delta_a$ as they are universal geometric dependence independent of the tensor network.

This result is amazing! Firstly, since $k,b,c$ do not depend on $\Delta_a$, there may be no solution to match (\ref{m1}) with (\ref{m2}) if (\ref{m1}) were not satisfied. The existence of the solution is a strong evidence that Einstein equation is indeed encoded in the tensor network. Secondly, supposing $k,b,c$ are known at the beginning, the mass-dimension relation can be derived from the Einstein equation.
Let us emphasize that it would be inconsistent had (\ref{m2}) not been satisfied. The reason is that the effective action that we wrote down to match the correlation functions read off from the tensor network already suggested that (\ref{m2}) is satisfied. Therefore the Einstein equation that descends from this effective action is only consistent if (\ref{m2}) is satisfied -- otherwise it would suggest that the tensor network does not admit an effective field theory description in the bulk.

We discover a known relation with a new perspective. Thirdly, $2b/c=-p$ means that to leading order
\begin{eqnarray}
\label{boxj}
  G\equiv\frac{\delta S_{EH}}{\delta d_{xy}} =  4b j_{xy}+c(\sum_{\substack{i\\(y_i\neq y)}} j_{xy_i}+\sum_{\substack{i\\(x_i\neq x)}}j_{x_iy}) = -c\Box j_{xy},
\end{eqnarray}
where the graph Laplacian for an edge degree of freedom $\Box j_e$ is defined by
\begin{eqnarray}
  \Box j_e &\equiv& \sum_{f\sim e}(j_e-j_f).
\end{eqnarray}
Here $\sum_{f\sim e}$ denotes the sum over all edges $f$ that share a vertex with a fixed edge $e$. Our result is similar to the one that appears in \cite{Gubser:2016htz}. The authors in \cite{Gubser:2016htz} start with a reasonable definition of Ricci curvature on graphs and then obtain a result similar to (\ref{boxj}).
Assuming only locality and isotopy in our ansatz and requiring that the resultant Einstein equation is satisfied leads to a unique expression that matches with the mathematics literature for $G$ in the perturbative limit away from the pure BT geometry.

Having $k=1, 2b/c=-p$, we further get
\begin{eqnarray}
  B^{aa} = \frac{1}{2 c (p+1) \left(1-p^{2 \Delta_a}\right)}.
\end{eqnarray}

\subsection{Solving the Einstein constraint to order $\lambda^3$}

To order $\mathcal{O}(\lambda^2)$, we have no restriction on the interaction term $\tilde{\phi}^3$ in the action since $\tilde{\phi}^3\sim \mathcal{O}(\lambda^3)$. While a solution for Einstein equation to this order is non-trivial -- as we have seen above it involved various non-trivial constraints on the mass that happens to be consistent with the AdS/CFT dictionary --  we note that the variation of the relative entropy also led to an Einstein equation to this order.
In this section we would show that such constraints can also be solved in the $\lambda^3$ order leading to a self-consistent result. This is certainly beyond kinematics and provide strong evidence that the correct dynamics are indeed encoded in the tensor network.

Now let's work out $G$ and $T$ to order $\mathcal{O}(\lambda^3)$ and see what kinds of interaction terms we will get. The procedure is the same as before, but the calculation becomes more complicated.

We still consider the graph shown in Fig. \ref{part}. The setup is the same as before. Now we have
\begin{eqnarray}
\label{phixa1}
  \tilde{\phi}_x^a =\lambda^a p^{-\Delta_a}+\tilde{\lambda}^a p^{-2\Delta_a}+\eta^a p^{-\Delta_a}+\tilde{\eta}^a p^{-2\Delta_a}+\gamma^a+\tilde{\gamma}^a p^{-\Delta_a}+\lambda^b\tilde{\lambda}^c p^{-\Delta_b}p^{-2\Delta_c}C^{abc}+\mathcal{O}(\lambda^3),\;\;\\
  \label{phiya1}
  \tilde{\phi}_y^a =\tilde{\lambda}^a p^{-\Delta_a}+\lambda^a p^{-2\Delta_a}+\tilde{\eta}^a p^{-\Delta_a}+\eta^a p^{-2\Delta_a}+\tilde{\gamma}^a+\gamma^a p^{-\Delta_a}+\tilde{\lambda}^b\lambda^c p^{-\Delta_b}p^{-2\Delta_c}C^{abc}+\mathcal{O}(\lambda^3),\;\;
\end{eqnarray}
where
\begin{eqnarray}
  \eta^a &\equiv& \sum_{i} \eta^a_i,\\
  \tilde{\eta}^a &\equiv& \sum_{i} \tilde{\eta}^a_i,\\
  \gamma^a&\equiv& \sum_{i\neq j}\lambda^b_i \lambda^c_j C^{abc}p^{-\Delta_b}p^{-\Delta_c},\\
  \tilde{\gamma}^a&\equiv& \sum_{i\neq j}\tilde{\lambda}^b_i \tilde{\lambda}^c_j C^{abc}p^{-\Delta_b}p^{-\Delta_c}.
\end{eqnarray}
Here $\sum_{i\neq j}$ denotes the sum of all possible ways of picking two edges from $p$ edges. We can also express $\gamma^a,\tilde{\gamma}^a$ as
\begin{eqnarray}
\label{gamma}
  \gamma^a &=& \frac{1}{2}(\lambda^b\lambda^c C^{abc}p^{-\Delta_b}p^{-\Delta_c}-\sum_k \lambda_k^b\lambda_k^c C^{abc}p^{-\Delta_b}p^{-\Delta_c}),\\
  \label{gammat}
 \tilde{\gamma}^a &=& \frac{1}{2}(\tilde{\lambda}^b\tilde{\lambda}^c C^{abc}p^{-\Delta_b}p^{-\Delta_c}-\sum_k \tilde{\lambda}_k^b\tilde{\lambda}_k^c C^{abc}p^{-\Delta_b}p^{-\Delta_c}).
\end{eqnarray}
Plugging (\ref{phixa1}), (\ref{phiya1}) into (\ref{T}), we get
\begin{tiny}
\begin{eqnarray}
\nonumber
   T_{\lambda^3} &=&  r_1 R^{abc} p^{-2 \left(\Delta _a+\Delta _b+\Delta _c\right)}  \left(\left(\tilde{\lambda }^a p^{\Delta _a}+\lambda ^a\right) \left(\tilde{\lambda }^b p^{\Delta _b}+\lambda ^b\right) \left(\tilde{\lambda }^c+\lambda ^c p^{\Delta _c}\right)+\left(\tilde{\lambda }^a+\lambda ^a p^{\Delta _a}\right) \left(\tilde{\lambda }^b+\lambda ^b p^{\Delta _b}\right) \left(\tilde{\lambda }^c p^{\Delta _c}+\lambda ^c\right)\right)\\
   \nonumber
   &\;&+ h_1 H^{abc} p^{-2 \left(\Delta _a+\Delta _b+\Delta _c\right)} \left(\left(\tilde{\lambda }^a+\lambda ^a p^{\Delta _a}\right) \left(\tilde{\lambda }^b+\lambda ^b p^{\Delta _b}\right) \left(\tilde{\lambda }^c+\lambda ^c p^{\Delta _c}\right)+\left(\tilde{\lambda }^a p^{\Delta _a}+\lambda ^a\right) \left(\tilde{\lambda }^b p^{\Delta _b}+\lambda ^b\right) \left(\tilde{\lambda }^c p^{\Delta _c}+\lambda ^c\right)\right)\\
   \nonumber
   &\;&+\frac{\left(\lambda ^a-\tilde{\lambda }^a\right) p^{-3 \Delta _a} \left(p^{\Delta _a}-1\right) \left(\left(p^{\Delta _a}-1\right) p^{-2 \Delta _a} \left(-\tilde{\eta }^a+\left(\gamma ^a-\tilde{\gamma }^a\right) p^{\Delta _a}+\eta ^a\right)+C^{abc} \left(\lambda ^b \tilde{\lambda }^c-\lambda ^c \tilde{\lambda }^b\right) p^{-\Delta _b-2 \Delta _c}\right)}{1-p^{-2 \Delta _a}}\\
   \nonumber
   &\;&+\frac{\left(p^{\Delta _a}-p\right) p^{-4 \Delta _a-\Delta _b-2 \Delta _c}}{(p+1) \left(p^{\Delta _a}+1\right)}\Big[\left(\tilde{\gamma }^a \tilde{\lambda }^a+2 \lambda ^a \tilde{\eta }^a+2 \eta ^a \tilde{\lambda }^a+\gamma ^a \lambda ^a\right) p^{\Delta _a+\Delta _b+2 \Delta _c}+\left(\tilde{\gamma }^a \tilde{\lambda }^a+\gamma ^a \lambda ^a\right) p^{3 \Delta _a+\Delta _b+2 \Delta _c}\\
   \nonumber
   &\;&+\left(2 \lambda ^a \tilde{\gamma }^a+2 \gamma ^a \tilde{\lambda }^a+\tilde{\eta }^a \tilde{\lambda }^a+\eta ^a \lambda ^a\right) p^{2 \Delta _a+\Delta _b+2 \Delta _c}+\left(\tilde{\eta }^a \tilde{\lambda }^a+\eta ^a \lambda ^a\right) p^{\Delta _b+2 \Delta _c}+C^{abc} p^{2 \Delta _a} \left(\lambda ^a \lambda ^c \tilde{\lambda }^b+\lambda ^b \tilde{\lambda }^a \tilde{\lambda }^c\right)\\
   &\;&+C^{abc} p^{3 \Delta _a} \left(\lambda ^c \tilde{\lambda }^a \tilde{\lambda }^b+\lambda ^a \lambda ^b \tilde{\lambda }^c\right)\Big].
\end{eqnarray}
\end{tiny}
Here $T_{\lambda^3}$ is the $\mathcal{O}(\lambda^3)$ term in $T$.

For edge $xy_i$, after careful analysis its $V^a$ and $\tilde{V}^a$ are
\begin{eqnarray}
  V^a &=& V^a_i=\delta_1^a+\lambda_i^a+\eta_i^a+\mathcal{O}(\lambda^3),\\
  \nonumber
  \tilde{V}^a &=&\delta_1^a+(\lambda^a-\lambda^a_i)p^{-\Delta_a}+\tilde{\lambda}^ap^{-2\Delta_a}
  +(\eta^a-\eta^a_i)p^{-\Delta_a}+\tilde{\eta}^ap^{-2\Delta_a}+\tilde{\gamma}^a p^{-\Delta_a}\\
  &\;&+(\lambda^b-\lambda^b_i)\tilde{\lambda}^c C^{abc}p^{-\Delta_b}p^{-2\Delta_c}
  +\gamma^a-\lambda^b_i(\lambda^c-\lambda^c_i)C^{abc}p^{-\Delta_b}p^{-\Delta_c}+\mathcal{O}(\lambda^3).
\end{eqnarray}
Similarly for edge $x_iy$, its $V^a$ and $\tilde{V}^a$ are
\begin{eqnarray}
  \tilde{V}^a &=& \tilde{V}^a_i=\delta_1^a+\tilde{\lambda}_i^a+\tilde{\eta}_i^a+\mathcal{O}(\lambda^3),\\
  \nonumber
  V^a &=&\delta_1^a+(\tilde{\lambda}^a-\tilde{\lambda}^a_i)p^{-\Delta_a}+\lambda^ap^{-2\Delta_a}
  +(\tilde{\eta}^a-\tilde{\eta}^a_i)p^{-\Delta_a}+\eta^ap^{-2\Delta_a}+\gamma^a p^{-\Delta_a}\\
  &\;&+(\tilde{\lambda}^b-\tilde{\lambda}^b_i)\lambda^c C^{abc}p^{-\Delta_b}p^{-2\Delta_c}
  +\tilde{\gamma}^a-\tilde{\lambda}^b_i(\tilde{\lambda}^c-\tilde{\lambda}^c_i)C^{abc}p^{-\Delta_b}p^{-\Delta_c}+\mathcal{O}(\lambda^3).
\end{eqnarray}
For edge $xy$, its $V^a$ and $\tilde{V}^a$ are
\begin{eqnarray}
  V^a &=& \delta_1^a+\lambda^a p^{-\Delta_a}+\eta^a p^{-\Delta_a}+\gamma^a+\mathcal{O}(\lambda^3),\\
\tilde{V}^a&=&\delta_1^a+\tilde{\lambda}^ap^{-\Delta_a}+\tilde{\eta}^a p^{-\Delta_a}+\tilde{\gamma}^a+\mathcal{O}(\lambda^3).
\end{eqnarray}

This time the precision of (\ref{de}) is not enough since it is only of order $\mathcal{O}(\lambda^2)$. Now we persevere to recover the order $\mathcal{O}(\lambda^3)$ term, which is given by
\begin{eqnarray}
\label{de2}
  d_e &=&1+B^{ab}(\omega_e^a\omega_e^b+\tilde{\omega}_e^a\tilde{\omega}_e^b)+C^{ab}\omega_e^a\tilde{\omega}_e^b
  +D^{abc}(\omega_e^a\omega_e^b\omega_e^c+\tilde{\omega}_e^a\tilde{\omega}_e^b\tilde{\omega}_e^c)
  +E^{abc}(\omega_e^a\omega_e^b\tilde{\omega}_e^c+\tilde{\omega}_e^a\tilde{\omega}_e^b\omega_e^c)+\mathcal{O}(\omega^4)  \nonumber\\
  \nonumber
  &=& 1+B^{ab}(\lambda_e^a\lambda_e^b+\tilde{\lambda}_e^a\tilde{\lambda}_e^b)
  +C^{ab}\lambda_e^a\tilde{\lambda}_e^b
  +2B^{ab}(\lambda_e^a\eta_e^b+\tilde{\lambda}_e^a\tilde{\eta}_e^b)
  +C^{ab}(\lambda_e^a\tilde{\eta}_e^b+\eta_e^a\tilde{\lambda}_e^b)\\
  &\;&+D^{abc}(\lambda_e^a\lambda_e^b\lambda_e^c+\tilde{\lambda}_e^a\tilde{\lambda}_e^b\tilde{\lambda}_e^c)
  +E^{abc}(\lambda_e^a\lambda_e^b\tilde{\lambda}_e^c+\tilde{\lambda}_e^a\tilde{\lambda}_e^b\lambda_e^c)+\mathcal{O}(\lambda^4),
\end{eqnarray}
where $D^{abc}$ is fully symmetric and $E^{abc}$ is symmetric in $a,b$. Here we have used $A^a=0$, which means expanding $V^a,\tilde{V}^a$ to $\mathcal{O}(\lambda^2)$ is enough to obtain $d_e$ to $\mathcal{O}(\lambda^3)$.

Plugging $V^a,\tilde{V}^a$ above into (\ref{de2}), we can obtain $j_{xy},\{j_{xy_i}\},\{j_{x_iy}\}$ to order $\mathcal{O}(\lambda^3)$. Plugging all the $j$ into (\ref{G}), we can obtain $G$ to order $\mathcal{O}(\lambda^3)$. Since $j\sim \mathcal{O}(\lambda^2)$, we have $j^2\sim \mathcal{O}(\lambda^4)$. Therefore it is consistent to ignore the $\mathcal{O}(j^2)$ term in $G$.

Let us denote the order $\mathcal{O}(\lambda^3)$ term of $G$ by $G_{\lambda^3}$. After careful calculation, we can work out  $G_{\lambda^3}+T_{\lambda^3}$. After using the $B^{ab},C^{ab},b=-pc/2$ we have obtained before, in $G_{\lambda^3}+T_{\lambda^3}$ the term which depends on $\{\eta^a_i\},\{\tilde{\eta}^a_i\}$ automatically vanishes. Finally we find
\begin{eqnarray}
\nonumber
  G_{\lambda^3}+T_{\lambda^3}&=& \sum_{j=1}^p M_1^{abc} (\lambda_j^a\lambda_j^b\lambda_j^c+\tilde{\lambda}_j^a\tilde{\lambda}_j^b\tilde{\lambda}_j^c)
  +\sum_{k=1}^p M_2^{abc}\left(\lambda^a_k\lambda^b_k(p^{\Delta_c}\lambda^c+\tilde{\lambda}^c)+\tilde{\lambda}^a_k\tilde{\lambda}^b_k(p^{\Delta_c}\tilde{\lambda}^c+\lambda^c)\right)\\
  &\;&+M_3^{abc}(\lambda^a\lambda^b\lambda^c+\tilde{\lambda}^a\tilde{\lambda}^b\tilde{\lambda}^c)
  +M_4^{abc}(\lambda^a\lambda^b\tilde{\lambda}^c+\tilde{\lambda}^a\tilde{\lambda}^b\lambda^c),
\end{eqnarray}
where
\begin{small}
\begin{eqnarray}
  M_1^{abc} &=& \frac{p^{-\Delta _a-\Delta _b-\Delta _c} \left(2 c (p+1) \left(D^{abc} \left(p^{\Delta _a+\Delta _b+\Delta _c}-1\right)+E^{abc} \left(p^{\Delta _c}-p^{\Delta _a+\Delta _b}\right)\right)-C^{abc}\right)}{2 (p+1)},\\
  M_2^{abc}&=&\frac{1}{4} p^{-\Delta _a-\Delta _b-2 \Delta _c} \left(4 c \left(E^{abc} p^{\Delta _a+\Delta _b}-E^{bca} p^{\Delta _a}-E^{cab} p^{\Delta _b}+3 D^{abc}\right)+\frac{C^{abc}}{\frac{4 (p+1)}{p^{2 \Delta _c}-5}+p+1}\right),\\
  \nonumber
  M_3^{abc}&=&\frac{1}{4} p^{-2 \left(\Delta _a+\Delta _b+\Delta_c\right)}\Bigg( 4 r_1 R^{abc} \left(p^{\Delta _a+\Delta _b}+p^{\Delta _c}\right)+4 h_1 H^{abc} \left(p^{\Delta _a+\Delta _b+\Delta _c}+1\right)+4 c E^{abc} p^{\Delta _a+\Delta _b+2 \Delta _c}\\
  &\;&+4 c D^{abc} \left(p-(p+3) p^{\Delta _a+\Delta _b+\Delta _c}\right)+\frac{C^{abc} \left(p^{2 \Delta _a}+3\right) p^{\Delta _a+\Delta _b+\Delta _c}}{(p+1) \left(p^{2 \Delta _a}-1\right)}\Bigg),\\
  \nonumber
  M_4^{abc}&=&\frac{1}{4} p^{-2 \left(\Delta _a+\Delta _b+\Delta _c\right)}\Bigg(12 c D^{abc} \left((p-2) p^{\Delta _a+\Delta _b}+(p-1) p^{\Delta _c}\right)+\\
  \nonumber
  &\;&C^{abc}\left((p+1) \left(p^{2 \Delta _a}-1\right) \left(p^{2 \Delta _c}-1\right)\right)^{-1}\Big[4 \left(p^{2 \Delta _a+1}-1\right) p^{\Delta _c}-4 \left(p^{2 \Delta _a+1}-1\right) p^{3 \Delta _c}\\
  \nonumber
  &\;&+\left(3 p^{2 \Delta _a}-4 p+1\right) p^{\Delta _a+\Delta _b+2 \Delta _c}+\left(p^{2 \Delta _a}+4 p-5\right) p^{\Delta _a+\Delta _b}\Big]+4\Big[3 h_1 H^{abc} p^{\Delta _a+\Delta _b}\\
  \nonumber
  &\;&+c E^{abc} p^{\Delta _c} \left(p^{\Delta _c}-2 p^{\Delta _a+\Delta _b+1}\right)+c E^{bca} p^{2 \Delta _a+\Delta _b}+c E^{cab} p^{\Delta _a+2 \Delta _b}+r_1 R^{cab} p^{\Delta _a}\\
  &\;&+r_1 R^{bca} p^{\Delta _b}+r_1 R^{abc}+p^{\Delta _c} \left(r_1 p^{\Delta _b} \left(R^{abc} p^{\Delta _a}+R^{cab}\right)+r_1 R^{bca} p^{\Delta _a}+3 h_1 H^{abc}\right)\Big]\Bigg).
\end{eqnarray}
\end{small}
Here we have used (\ref{gamma}),(\ref{gammat}), so $\gamma^a,\tilde{\gamma}^a$ do not appear in our final expression.

It is obvious that the Einstein equation is
\begin{eqnarray}
  G_{\lambda^3}+T_{\lambda^3} &=& 0,
\end{eqnarray}
for any $\{\lambda^a_i\},\{\tilde{\lambda}^a_i\}$. So we should have
\begin{eqnarray}
\label{Ms}
  M_1^{(abc)} = 0,\;\;
  M_2^{(ab)c} = 0,\;\;
  M_3^{(abc)} = 0,\;\;
  M_4^{(ab)c} = 0,
\end{eqnarray}
where $(ab),(abc)$ means the symmetrization of the matrix. The unknown matrices  are $D^{abc}=D^{(abc)},E^{abc}=E^{(ab)c},H^{abc}=H^{(abc)},R^{abc}=R^{(ab)c}$.
They can be fully determined by (\ref{Ms}) since the number of unknown independent matrix elements is the same as the number of independent equations.

Using $M_1^{(abc)} = 0$, we can get
\begin{small}
\begin{eqnarray}
\label{Dabc}
  D^{abc} =\frac{3C^{abc}+2c(1+p)\left((p^{\Delta_a+\Delta_b}-p^{\Delta_c})E^{abc}
  +(p^{\Delta_b+\Delta_c}-p^{\Delta_a})E^{bca}
  +(p^{\Delta_a+\Delta_c}-p^{\Delta_b})E^{cab}\right)}{6c(1+p)(-1+p^{\Delta_a+\Delta_b+\Delta_c})}.
\end{eqnarray}
\end{small}
Plugging it into $M_2^{(ab)c} = 0$, we can get
\begin{small}
\begin{eqnarray}
 \nonumber
  &\;&(p^{2 (\Delta_a+\Delta_b)}-1) E^{abc}-\left(p^{2 \Delta_a}-1\right) p^{\Delta_b} E^{bca}+p^{\Delta_a} \left(-\left(p^{2 \Delta_b}-1\right)\right) E^{cab}\\
  &\;&=-\frac{C^{abc}p^{-\Delta_c} \left(-5 p^{\Delta_a+\Delta_b+\Delta_c}+p^{\Delta_a+\Delta_b+3 \Delta_c}+5 p^{2 \Delta_c}-1\right)}{4 c (p+1) \left(p^{2 \Delta_c}-1\right)}.
\end{eqnarray}
\end{small}
By solving it, we obtain
\begin{small}
\begin{eqnarray}
\label{Eabc}
  E^{abc}=\frac{C^{abc}( -p^{\Delta _a+\Delta _b+2 \Delta _c}-p^{\Delta _a-\Delta _b}-p^{\Delta _b-\Delta _a}+3 p^{\Delta _a+\Delta _b}-p^{2 \Delta _a+\Delta _c}-p^{2 \Delta _b+\Delta _c}-p^{-\Delta _c}+3 p^{\Delta _c})}{4 c (p+1) \left(p^{2 \Delta _a}-1\right) \left(p^{2 \Delta _b}-1\right) \left(p^{2 \Delta _c}-1\right)}.
\end{eqnarray}
\end{small}

Using $M_3^{(abc)} = 0$, we can get
\begin{small}
\begin{eqnarray}
\nonumber
\label{Habc}
  H^{abc} &=& -\frac{1}{12 h_1 \left(p^{\Delta _a+\Delta _b+\Delta _c}+1\right)}\Big(\frac{p^{\Delta _a+\Delta _b+\Delta _c}C^{abc}}{(p+1) \left(p^{2 \Delta _a}-1\right) \left(p^{2 \Delta _b}-1\right) \left(p^{2 \Delta _c}-1\right)}\times\\
  \nonumber
  &\;&\left(3 p^{2 \Delta _a+2 \Delta _b+2 \Delta _c}+p^{2 \Delta _a+2 \Delta _b}+p^{2 \Delta _a+2 \Delta _c}+p^{2 \Delta _b+2 \Delta _c}-5 p^{2 \Delta _a}-5 p^{2 \Delta _b}-5 p^{2 \Delta _c}+9\right)\\
  \nonumber
  &\;&+4\big[3 c D^{abc} \left(p-(p+3) p^{\Delta _a+\Delta _b+\Delta _c}\right)+c E^{abc} p^{\Delta _a+\Delta _b+2 \Delta _c}+c E^{bca} p^{2 \Delta _a+\Delta _b+\Delta _c}+c E^{cab} p^{\Delta _a+2 \Delta _b+\Delta _c}\\
  &\;&+r_1 R^{abc} p^{\Delta _a+\Delta _b}+r_1 R^{cab} p^{\Delta _a+\Delta _c}+r_1 R^{bca} p^{\Delta _b+\Delta _c}+r_1 R^{bca} p^{\Delta _a}+r_1 R^{cab} p^{\Delta _b}+r_1 R^{abc} p^{\Delta _c}\big]\Big).
\end{eqnarray}
\end{small}
Plugging (\ref{Dabc}),(\ref{Eabc}),(\ref{Habc}) into $M_4^{(ab)c} = 0$, we can get
\begin{eqnarray}
  2 R^{abc} \left(p^{2 \left(\Delta _a+\Delta _b\right)}-1\right)+2 R^{bca} \left(p^{2 \Delta _a}-1\right) p^{\Delta _b}+2 R^{cab} p^{\Delta _a} \left(p^{2 \Delta _b}-1\right) = \frac{C^{abc}p^{\Delta _a+\Delta _b}}{r_1}.
\end{eqnarray}
By solving it, we obtain
\begin{eqnarray}
\label{Rabc1}
  R^{abc} &=& \frac{C^{abc}\left(p^{\Delta _a+\Delta _c}-p^{\Delta _b}\right) \left(p^{\Delta _b+\Delta _c}-p^{\Delta _a}\right)}{2 r_1 \left(p^{2 \Delta _a}-1\right) \left(p^{2 \Delta _b}-1\right) \left(p^{2 \Delta _c}-1\right)}.
\end{eqnarray}

Having $E^{abc},R^{abc}$, we can further get
\begin{small}
\begin{eqnarray}
\nonumber
  D^{abc}&=&\frac{-C^{abc}p^{-\Delta _a-\Delta _b-\Delta _c}}{12 c (p+1) \left(p^{2 \Delta _a}-1\right) \left(p^{2 \Delta _b}-1\right) \left(p^{2 \Delta _c}-1\right)}\Big(-3 p^{\Delta _a+\Delta _b+\Delta _c}-3 p^{2 \left(\Delta _a+\Delta _b+\Delta _c\right)}\\
  &\;&+p^{3 \Delta _a+\Delta _b+\Delta _c}+p^{\Delta _a+3 \Delta _b+\Delta _c}+p^{\Delta _a+\Delta _b+3 \Delta _c}+p^{2 \left(\Delta _a+\Delta _b\right)}+p^{2 \left(\Delta _a+\Delta _c\right)}+p^{2 \left(\Delta _b+\Delta _c\right)}\Big),
  \label{Dabc1}\\
  \nonumber
  H^{abc}&=&\frac{C^{abc}p^{-\Delta _a-\Delta _b-\Delta _c} \left(p^{\Delta _a+\Delta _b+\Delta _c}+p\right)}{12 h_1 (p+1) \left(p^{2 \Delta _a}-1\right) \left(p^{2 \Delta _b}-1\right) \left(p^{2 \Delta _c}-1\right)}\Big(-3 p^{\Delta _a+\Delta _b+\Delta _c}-3 p^{2 \left(\Delta _a+\Delta _b+\Delta _c\right)}\\
  &\;&+p^{3 \Delta _a+\Delta _b+\Delta _c}+p^{\Delta _a+3 \Delta _b+\Delta _c}+p^{\Delta _a+\Delta _b+3 \Delta _c}+p^{2 \left(\Delta _a+\Delta _b\right)}+p^{2 \left(\Delta _a+\Delta _c\right)}+p^{2 \left(\Delta _b+\Delta _c\right)}\Big).
  \label{Habc1}
\end{eqnarray}
\end{small}
As we can see, $R^{abc}$ is simple, while $E^{abc},D^{abc},H^{abc}$ are a bit complicated. And we notice that
\begin{eqnarray}
  \frac{D^{abc}}{H^{abc}} &=& \frac{-h_1}{c(p^{\Delta_a+\Delta_b+\Delta_c}+p)}.
\end{eqnarray}

It is not trivial that $D^{abc},E^{abc},R^{abc},H^{abc}$ can be exactly solved. It means that to satisfy the Einstein equation the interaction term and the definition of edge length are fixed!

\subsubsection{The interaction term in the semi-classical limit}
As we have noted in section 2, the field theory exactly recovers the correlation functions in the tensor network in the semi-classical limit where the masses approach infinity. Therefore we would like to take this limit in the results obtained in the previous section.

Recall that $\langle \tilde{\phi}^a_x\tilde{\phi}^b_y\tilde{\phi}^c_z\rangle =  \langle O^a_xO^b_yO^c_z\rangle$.
This is consistent with an action with cubic term.
\begin{eqnarray}
\label{actioni}
S_{cubic} =  \frac{1}{3!}  \sum_x C^{abc}\tilde{\phi}^a_x\tilde{\phi}^b_x\tilde{\phi}^c_x.
\end{eqnarray}
At first glance our $H^{abc},R^{abc}$ do not agree with (\ref{actioni}).
However, when we take the classical limit, i.e. $\Delta \gg 1$, we have
\begin{eqnarray}
  H^{abc} &=& \frac{-C^{abc}}{4h_1(1+p)},\\
  R^{abc}&=&0,
\end{eqnarray}
where we have used $\Delta_a\sim\Delta_b\sim\Delta_c\gg1$. Then the interaction term in the action (\ref{actionm}) becomes
\begin{eqnarray}
  \sum_{\langle xy\rangle}\frac{-h_0C^{abc}}{4h_1(1+p)}(\tilde{\phi}^a_x\tilde{\phi}^b_x\tilde{\phi}^c_x+\tilde{\phi}^a_y\tilde{\phi}^b_y\tilde{\phi}^c_y)
  =\sum_{x}\frac{-h_0C^{abc}}{4h_1}\tilde{\phi}^a_x\tilde{\phi}^b_x\tilde{\phi}^c_x.  \label{eq:actioni_Einstein}
\end{eqnarray}
Here we consider the vacuum case so that $h(d_{xy})=h_0$ and we have changed the sum over edges into vertices.
We have noted in section 2 that at every vertex where propagators meet, the tensor network dictates that it is weighted by a factor of the fusion
coefficients. In particular when there are exactly three propagators meeting at the vertex, the interaction coupling is precisely $C^{abc}$.

Up to some $\Delta_a$ independent normalizations that is not fixed by the Einstein constraints, the latter led to (\ref{eq:actioni_Einstein}) that is in complete agreement with results consistent with the tensor network.

\subsubsection{Fisher metric and the edge lengths}
In the definition of $d_e$, $D^{abc},E^{abc}$ are a bit complicated. We find a way to simplify the expression of $d_e$. Originally it is given by
\begin{eqnarray}
  \nonumber
  d_e
  &=& 1+B^{ab}(\lambda_e^a\lambda_e^b+\tilde{\lambda}_e^a\tilde{\lambda}_e^b)
  +C^{ab}\lambda_e^a\tilde{\lambda}_e^b
  +2B^{ab}(\lambda_e^a\eta_e^b+\tilde{\lambda}_e^a\tilde{\eta}_e^b)
  +C^{ab}(\lambda_e^a\tilde{\eta}_e^b+\eta_e^a\tilde{\lambda}_e^b)\\
  &\;&+D^{abc}(\lambda_e^a\lambda_e^b\lambda_e^c+\tilde{\lambda}_e^a\tilde{\lambda}_e^b\tilde{\lambda}_e^c)
  +E^{abc}(\lambda_e^a\lambda_e^b\tilde{\lambda}_e^c+\tilde{\lambda}_e^a\tilde{\lambda}_e^b\lambda_e^c)+\mathcal{O}(\lambda^4).
\end{eqnarray}
Supposing the two ends of edge $e$ are $u,v$, $V^a$ is on $u$ side and $\tilde{V}^a$ is on $v$ side. Then
\begin{eqnarray}
  \tilde{\phi}_u^a &=& \lambda_e^a+\tilde{\lambda}_e^a p^{-\Delta_a}+\eta^a_e+\tilde{\eta}_e^a p^{-\Delta_a}+C^{abc}\lambda^b\tilde{\lambda}^cp^{-\Delta_c}+\mathcal{O}(\lambda^3),\\
  \tilde{\phi}_v^a &=& \tilde{\lambda}_e^a+\lambda_e^a p^{-\Delta_a}+\tilde{\eta}^a_e+\eta_e^a p^{-\Delta_a}+C^{abc}\tilde{\lambda}^b\lambda^cp^{-\Delta_c}+\mathcal{O}(\lambda^3).
\end{eqnarray}
We find that to order $\mathcal{O}(\lambda^3)$
\begin{eqnarray}
  1+\frac{p^{\Delta _a}}{2 c (p+1) \left(1-p^{2 \Delta _a}\right)}\tilde{\phi}^a_u\tilde{\phi}^a_v +\frac{ r_1 R^{abc}}{2c(p+1)}(\tilde{\phi}^a_u\tilde{\phi}^b_u\tilde{\phi}^c_v+\tilde{\phi}^a_v\tilde{\phi}^b_v\tilde{\phi}^c_u)= d_e.
\end{eqnarray}

The nonlocal term in $T$ is given by
\begin{eqnarray}
\nonumber
  T^{\diamondsuit}_{xy} &=& \frac{-2\zeta_p(2\Delta_a)}{2p^{\Delta_a}}\tilde{\phi}^a_x\tilde{\phi}^a_y+
  r_1R^{abc}(\tilde{\phi}^a_x\tilde{\phi}^b_x\tilde{\phi}^c_y+\tilde{\phi}^a_y\tilde{\phi}^b_y\tilde{\phi}^c_x)\\
  &=&\frac{p^{\Delta_a}}{1-p^{2\Delta_a}}\tilde{\phi}^a_x\tilde{\phi}^a_y+
  r_1R^{abc}(\tilde{\phi}^a_x\tilde{\phi}^b_x\tilde{\phi}^c_y+\tilde{\phi}^a_y\tilde{\phi}^b_y\tilde{\phi}^c_x).
\end{eqnarray}
So it turns out that
\begin{eqnarray}
  1+\frac{T^\diamondsuit_{uv}}{2c(p+1)} &=& d_{e}.
\end{eqnarray}
The edge length is closely related to the nonlocal part of $T$.

We can also introduce some states on the vertices $u,v$:
\begin{eqnarray}
  |u\rangle &\equiv& \sum_a \mathcal{N}_a(\tilde{\phi}^a_u+\tilde{\phi}^b_u\tilde{\phi}^c_u \tilde{R}^{bca} )|a\rangle,\\
  |v\rangle &\equiv& \sum_a  \mathcal{N}_a (\tilde{\phi}^a_v+\tilde{\phi}^b_v\tilde{\phi}^c_v \tilde{R}^{bca} )|a\rangle,
\end{eqnarray}
where
\begin{eqnarray}
 \mathcal{N}_a &=& \sqrt{\frac{p^{\Delta_a}}{2 c(p+1)(p^{2\Delta_a }-1)}}, \\
  \tilde{R}^{abc}&\equiv&\frac{r_1R^{abc}(1-p^{2\Delta_c}) }{p^{\Delta_c}}.
\end{eqnarray}
Then we have
\begin{eqnarray}
  1-\langle u|v\rangle &=& d_{e}.
\end{eqnarray}
So $d_e$ may also be understood as some information metric.
In particular, let us consider the semi-classical limit that is of direct relevance in the comparison with the tensor network.
One readily finds that
\be
\lim_{\Delta_a \to \infty} |u \rangle =  \frac{1}{\sqrt{2c (p+1)}}  \sum_{a} {\phi}^a_u |a\rangle.
\ee
The wavefunction of this state is obtained by sticking a dangling leg to the vertex $u$ of the tensor network.
The edge distance turns out to be given by the Fisher information distance between two states each with a leg stuck at one of the two ends of the edge!

\section{Summary and Discussion}
In this paper, we study in detail the tensor network proposed in \cite{Hung:2019zsk} with deformed boundary conditions. We show that there is virtually a unique way of assigning geometrical data to the tensor network based on its local data so that it satisfies a graph Einstein equation that is self-consistent with the field theory that is known to be encoded in the tensor network. The emergent Einstein tensor also naturally recovers the proposals in the mathematics literature, at least in the perturbative expansion that we are considering \cite{Yau1,Ollivier,Gubser:2016htz}.
This is arguably the first such quantitative demonstration of an emergent Einstein equation with matter in a holographic tensor network that explicitly reconstructs known CFTs in a controlled way, albeit (unfortunately) in a simplified setting of the p-adic AdS/CFT.

To summarise our feat slightly differently, the search for Einstein equation is to look for a question whose answer we have -- a tensor network carrying the CFT data. And we constructed such a relation guided by the AdS/CFT correspondence, which helped us in finding a bulk action that encodes the same correlation data as the tensor network.

There are several interesting observations and future problems that we would like to comment upon.
First of all, in the covariant tensor network reconstruction of the CFT partition function rather than a wavefunction, the same choices of tensors can describe different geometries. The important aspect is the sub-vector space that is explored -- which is controlled by the boundary conditions in the current construction. Near a fixed point vector, only vectors close to the fixed point vector is actually contributing to the partition function.
When a different sub-space is explored, the same tensor network can describe a vastly different geometry. This is in fact very close in spirit with the gravity path-integral, where different saddle points are merely exploring different vector subspaces.

Second, the tensor network construction here bears uncanny resemblance to strange correlators, a term coined in \cite{2014PhRvL.112x7202Y, Aasen:2020jwb, Bal:2018wbw,Aasen:2020jwb}. PEPS tensor networks produce  (minimal) CFT partition function by projecting each of the boundary leg to a particular fixed point vector.  Here, something almost identical happens. We can take the partition function we have constructed as the overlap of a direct product state and a MERA like tensor network state that covers the BT tree. i.e.
\be
Z_{CFT} = (\prod_i \langle V_f|)  | \Psi\rangle,
\ee
where $|\Psi\rangle $ is a ``state'' corresponding to the tensor network covering the BT tree and whose ``physical legs'' are the dangling ones at the cutoff surface. We are currently pushing this analogy to construct holographic tensor networks that describe real CFTs.
Moreover, we find our tensor network also a discrete realization of the recent consideration of \cite{Ma:2020pmt}, where the partition function of a generic QFT can always be thought of as the overlap between a ``fixed point'' state (which should be some direct product or gapped state) and another wavefunction in one higher dimensions. By considering RG transformation of the wavefunctions (which keeps the fixed point state invariant), one generates a holographic radial direction -- which is precisely what our BT tree tensor network achieved. Our construction of geometries and Einstein equations therefore could inspire more general constructions in real CFTs by making better use of this connection.

Our tensor network is also known to be equivalent to a Wilson line network with gauge group given by $SL(2,Q_p)$ \cite{Hung:2018mcn}. This is parallel to the Chern-Simons formulation of the 3d Einstein-Hilbert gravity \cite{Witten:1988hc}.
Recently, it is noticed that Wilson line networks provide more direct link with complexities. It would be very interesting to translate these results here, which might give an alternative way of assigning curvatures to the tensor network.

Indeed given our tensor network's connection with Wilson lines, it inspired an alternative way of bending the BT tree -- by deforming the connection upon which the Wilson line network is evaluated. In the process, one discovers a formulation of the BTZ black hole in the p-adic Wilson line network that bears profound resemblance to the AdS BTZ black hole, but unique in its way. We have also found a new set of coordinates on the BT tree that is the analogue of black hole coordinates in AdS space and opens the door to more general diffeomorphism transformation on the p-adic tree. The p-adic CFT, known for its lack of descendants, should still contain intricate structures and accommodate local conformal transformation. These exciting developments will be discussed in detail in our sequel to appear together!

Before we end, we note that we have expanded only up to $\lambda^3$ in the current paper. Since the distance perturbation $j$ is quadratic in $\lambda$, we obtained only a linear equation on $j$. Starting from $\lambda^4$ contributions of $j^2$ would show up in the Einstein constraint. It would be interesting to find out if the tensor network describes a truly ``pure Einstein'' action, or if it involves other curvature terms.

\textit{Acknowledgements.---} LYH acknowledges the support of NSFC (Grant
No. 11922502, 11875111) and the Shanghai Municipal Science and Technology Major Project
(Shanghai Grant No.2019SHZDZX01), and Perimeter Institute for hospitality as a part of the Emmy Noether Fellowship programme.
Part of this work was instigated in KITP during the program qgravity20.  LC acknowledges support of NSFC (Grant No. 12047515).
We thank  Bartek Czech, Muxin Han, Ce Shen, Gabriel Wong, Qifeng Wu, Mathew Yu and Zhengcheng Gu for useful discussions and comments.
We thank Si-nong Liu and Jiaqi Lou for collaboration on related projects.

\appendix
\section{Brief review of p-adic CFTs}\label{padiccft}
Let's begin with the brief introduction of padic number field $Q_p$ which is the field extension of the rational numbers alternative to the real number $\mathds{R}$. For a given prime number $p$, any rational number $r\in \mathds{Q}$ can be written as
\begin{eqnarray}
  r &=& p^k \frac{a}{b}, \;\;\;(k\in \mathds{Z})
\end{eqnarray}
where $a$ and $b$ are integers relatively prime to $p$. And the p-adic norm of $r$ is defined as
\begin{eqnarray}
  |r|_p &=& p^{-k},\;\;\;\;|0|_p=0.
\end{eqnarray}
After the field extension by using the p-adic norm $|\dots|_p$, the resulting number field is called p-adic number field $Q_p$. A p-adic number $x\in Q_p$ is usually expressed
as an infinite series
\begin{eqnarray}
  x &=& p^v \sum^\infty_{i=0} a_i p^i, \;\;\;(v\in \mathds{Z})
\end{eqnarray}
where $a_i\in \mathds{F}_p=\{0,1,\dots,p-1\}$ and $a_0\neq 0$. And its p-adic norm is
\begin{eqnarray}
  |x|_p &=& p^{-v},
\end{eqnarray}
which satisfies various axioms of norms:
\begin{eqnarray}
  |x| &\geq& 0,\\
  |x|=0&\leftrightarrow& x=0,\\
  |xy|&=&|x||y|,\\
  |x+y|&\leq&|x|+|y|.
\end{eqnarray}
Actually the Euclidean norm and the p-adic norm are the only two types of norms that obey the four axioms above. Other than the real number $\mathds{R}$, the padic number $Q_p$ is the only possible field extension of rational number $\mathds{Q}$.

The padic CFT is a field theory living in the p-adic number field $Q_p$, i.e. its coordinates $x\in Q_p$. The global conformal symmetry of the padic CFT is defined as the transformation
\begin{eqnarray}
  x &\rightarrow& x^\prime=\frac{ax+b}{cx+d}, \;\;\;(a,b,c,d\in Q_p).
\end{eqnarray}
It furnishes the matrix group $\textrm{PGL}(2,Q_p)$, the direct analogue of $\textrm{SL}(2,\mathds{R})$ in $1d$ conformal transformation in real space-time. Similar to the CFT, there are two pieces of algebraic data required to specify completely a p-adic CFT:
\begin{itemize}
  \item First the spectrum of primary operators $\mathcal{O}_a$ with conformal dimensions $\Delta_a$, which transform under conformal symmetry $\textrm{PGL}(2,Q_p)$ as
      \begin{eqnarray}
        \mathcal{O}_a(x) &\rightarrow& \tilde{\mathcal{O}}_a(x^\prime)=\left|\frac{ad-bc}{(cx+d)^2}\right|_p^{-\Delta_a}\mathcal{O}_a(x),
      \end{eqnarray}
  \item Second, OPE coefficients $C^{ab}_{\;\;\;c}$ defined as
      \begin{eqnarray}
      \label{ope}
        \mathcal{O}_a(x_1)\mathcal{O}_b(x_2) &=& \sum_c C^{ab}_{\;\;\;c}|x_1-x_2|_p^{\Delta_c-\Delta_a-\Delta_b}\mathcal{O}_c(x_2).
      \end{eqnarray}
\end{itemize}
The form of (\ref{ope}) follows from the fact that the p-adic CFT has no descendant which is a consequence of the fact that on padic number field a locally-constant function has zero-derivative. As usual, there must be a unique identity operator $\mathds{I}=\mathcal{O}_1$ of dimension $\Delta_1=0$ whose OPE coefficients satisfy $C^{b1}_{\;\;\;a}=C^{1b}_{\;\;\;a}=\delta^b_a$.
The totally symmetric OPE coefficients $C^{abc}\equiv \sum_d C^{ab}_{\;\;\;d}C^{cd}_{\;\;\;1}$ is most commonly used. After suitable orthogonalization and normalization, we can always make $C^{ab}_{\;\;\;1}=\delta^{ab}$. Then $C^{1ab}=C^{a1b}=C^{ab1}=\delta^{ab}$. And inserting the OPE into a 2-point function implies that
\begin{eqnarray}
  \langle \mathcal{O}_a(x_1)\mathcal{O}_b(x_2)\rangle &=& \frac{\delta^{ab}}{|x_1-x_2|_p^{2\Delta_a}}.
\end{eqnarray}
For the 3-point function we will obtain
\begin{eqnarray}
  \langle \mathcal{O}_a(x_1)\mathcal{O}_b(x_2)\mathcal{O}_c(x_3)\rangle &=& \frac{C^{abc}}{|x_{12}|_p^{\Delta_a+\Delta_b-\Delta_c}
  |x_{23}|_p^{\Delta_b+\Delta_c-\Delta_a}
  |x_{31}|_p^{\Delta_c+\Delta_a-\Delta_b}}.
\end{eqnarray}
When considering the 4-point function, the crossing symmetry will lead to following constraint
\begin{eqnarray}
  \sum_c C^{abc}C^{cde} &=& \sum_c C^{bdc}C^{cae},
\end{eqnarray}
which is equivalent to the associativity of the fusion algebra defined by $C^{ab}_{\;\;\;c}$.

\section{Relation between cubic term and quadratic term}
To avoid confusion, the summation of $a,b,c$ in (\ref{T}) and (\ref{de2}) is understood as $\sum_{a\neq 1,b\neq 1,c\neq 1}$. Then our result is valid for $a\neq 1,b\neq 1,c\neq 1$. Now let's consider
\begin{small}
\begin{eqnarray}
\sum_{a,b,c} h_1H^{abc}(\tilde{\phi}^a_x\tilde{\phi}^b_x\tilde{\phi}^c_x+\tilde{\phi}^a_y\tilde{\phi}^b_y\tilde{\phi}^c_y)
+\sum_{a,b,c} r_1R^{abc}(\tilde{\phi}^a_x\tilde{\phi}^b_x\tilde{\phi}^c_y+\tilde{\phi}^a_y\tilde{\phi}^b_y\tilde{\phi}^c_x).
\end{eqnarray}
\end{small}
When one of $a,b,c$ is $1$, it will contribute
\begin{eqnarray}
\label{3to2}
\sum_{a\neq 1}\left( 3h_1H^{aa1}(\tilde{\phi}^a_x\tilde{\phi}^a_x+\tilde{\phi}^a_y\tilde{\phi}^a_y)
+ r_1R^{aa1}(\tilde{\phi}^a_x\tilde{\phi}^a_x+\tilde{\phi}^a_y\tilde{\phi}^a_y)
+2 r_1R^{1aa}(\tilde{\phi}^a_x\tilde{\phi}^a_y+\tilde{\phi}^a_y\tilde{\phi}^a_x)\right).
\end{eqnarray}
Here we have used $\tilde{\phi}^1_x=1+\dots$, $C^{ab1}=\delta^{ab}$ and the symmetry of $H^{abc},R^{abc}$. Plugging (\ref{Rabc1}),(\ref{Habc1}) into (\ref{3to2}), we will get
\begin{eqnarray}
  \sum_{a\neq 1}\left(
  -\frac{p^{2 \Delta _a}+p}{2 (p+1) \left(p^{2 \Delta _a}-1\right)}(\tilde{\phi}^a_x\tilde{\phi}^a_x+\tilde{\phi}^a_y\tilde{\phi}^a_y)
  +\frac{p^{\Delta _a}}{ p^{2 \Delta _a}-1}
  (\tilde{\phi}^a_x\tilde{\phi}^a_y)\right),
\end{eqnarray}
which is exactly
\begin{eqnarray}
  -\sum_{a\neq 1}\left( k\frac{\zeta_p(2\Delta_a)}{2p^{\Delta_a}}(\tilde{\phi}^a_x-\tilde{\phi}^a_y)^2+ \frac{\zeta_p(2\Delta_a)}{2p^{\Delta_a}}\frac{1}{p+1}m_a^2((\tilde{\phi}^{a}_x)^2+(\tilde{\phi}^{a}_y)^2) \right)=-T_{\tilde{\phi}^2}.
\end{eqnarray}
Here $T_{\tilde{\phi}^2}$ is the quadratic term in $T$ as shown in (\ref{T}). It implies that regarding the identity operator $\mathds{I}$ as an elementary field, in the action the quadratic term can be included in the interaction term in a compact way.

\section{Some identities in the flow of the boundary conditions}
As described in section 2.2, the authors of \cite{Hung:2019zsk} assume translation invariance along \({Q}_{p}\). Following (2.12), we also find some interesting relationship between the Fisher metric and the matter field.
Let the cutoff be denoted by \(\Lambda\), without normalization,
\begin{equation}
|V_{\Lambda}\rangle=(a_{\Lambda}+\sum_{k}b_{\Lambda}^{(k)}v^k)|1\rangle
\end{equation}
These are the boundary vectors, and p such vectors will return a new vector sharing the same vertex tensor with them, which is
\begin{equation}
|V_{\Lambda-1}\rangle=(a_{\Lambda-1}+\sum_{k}b_{\Lambda-1}^{(k)}v^k)|1\rangle
\end{equation}
in which \({v}^{k}\) represents some primary field.
According to 2.12, we have
\begin{equation}
|V_{\Lambda-1}\rangle=(a_{\Lambda}+\sum_{k}p^{-\Delta_k}b_{\Lambda}^{(k)}v^k)^{p}|1\rangle
\end{equation}
\begin{equation}
a_{\Lambda-1}=a_{\Lambda}^{p}+C_{p}^{2}\sum_{i,j}N_{ij}^{0}a_{\Lambda}^{p-2}b_{\Lambda}^{i}b_{\Lambda}^{j}p^{-\Delta_{i}-\Delta_{j}}+C_{p}^{3}\sum_{i,j,k,w}C^{wk0}C^{ijw}a_{\Lambda}^{p-3}b_{\Lambda}^{i}b_{\Lambda}^{j}b_{\Lambda}^{k}p^{-\Delta_{i}-\Delta_{j}-\Delta{k}}+\dots
\end{equation}
\begin{equation}
b_{\Lambda-1}^{(k)}=p^{1-\Delta_{k}}a_{\Lambda}^{p-1}b_{\Lambda}^{(k)}+C_{p}^{2}\sum_{i,j}C^{ijk}b_{\Lambda}^{(i)}b_{\Lambda}^{(j)}p^{-\Delta_{i}-\Delta_{j}}+\dots
\end{equation}

And
\begin{equation}
\lambda_k=\frac{b_\Lambda^{(k)}}{a_{\Lambda}}
\end{equation}

plays the role of a source, which satisfies,
\begin{equation}
\lambda_k << p^{\Delta_k-1}
\end{equation}

The fisher metric is
\begin{equation}
1-\frac{|\langle V_{\Lambda-n}|V_{\Lambda-n-1}\rangle|^2}{\langle V_{\Lambda-n}|V_{\Lambda-n}\rangle\langle V_{\Lambda-n-1}|V_{\Lambda-n-1}\rangle}
\end{equation}

Let n be the steps from the cutoff surface to the layer we are considering, and expand the fisher metric  between the nth layer and n+1th layer to first order, the recurrence relation tells us that the fisher metric should be

\begin{equation}
1-\frac{(a_{\Lambda-n-1}a_{\Lambda-n}+\sum_{k}b_{\Lambda-n-1}^{(k)}b_{\Lambda-n}^{(k)})^2}{(a_{\Lambda-n}^2+\sum_{k}{b_{\Lambda-n}^{(k)}}^2)(a_{\Lambda-n-1}^2+\sum_{k}{b_{\Lambda-n-1}^{(k)}}^2)}=\sum_{k}p^{2n-2n\Delta_{k}}(p^{1-\Delta_{k}}-1)^2\frac{{b_{\Lambda}^{(k)}}^{2}}{{a_{\Lambda}}^{2}}
\end{equation}

Now consider the expectation value of the operator inserted into some vertex x which is n steps from the boundary of the tree, to first order it is
\begin{equation}
\langle\phi_{x}\rangle=\sum_{r}JG(x,r)
\end{equation}

First ignore the normalization factor of the bulk-boundary propagator defined in \cite{Gubser:2016guj}, which is
\begin{equation}
C=\frac{1-p^{s-2\Delta_k}}{1-p^{-2\Delta_k}}
\end{equation}

Here s=1.

We have
\begin{equation}
\phi_{x}^{(k)}=[p^{n}p^{-n\Delta_k}+(p^{n+1}-p^{n})p^{-n\Delta_k-2\Delta_k})+\dots+(p^{n+w}-p^{n+w-1})p^{-n\Delta_k-2w\Delta_k}]\lambda^k
\end{equation}

Similarly,
\begin{equation}
\phi_{y}^{(k)}=[p^{n+1-(n+1)\Delta_k}(1-p^{-2\Delta_k})\frac{1-p^{(w-1)-2(w-1)\Delta_k}}{1-p^{1-2\Delta_k}}+p^{n+w}p^{-(n+1)\Delta_k-2(w-1)\Delta_k}]\lambda^k
\end{equation}

y is the point just below x, and w represents the distance between x and the IR limit, which
goes to infinity. When \({\Delta}_{k}\) is larger than \(\frac{1}{2}\),

\begin{equation}
(\phi_{x}^{(k)}-\phi_{y}^{(k)})^2=p^{2n-2n\Delta_k}(p^{1-\Delta_k}-1)^{2}(\frac{1-p^{-2\Delta_k}}{1-p^{1-2\Delta_k}})^{2}{\lambda^k}^{2}
\end{equation}

Now take the normalization factor into account and let

\begin{equation}
\Phi_{x}^{(k)}=C\phi_{x}^{(k)}
\end{equation}

We can see that \((\Phi_{x}^{(k)}-\Phi_{y}^{(k)})^{2}\) is just equal to the first order in the expansion of the Fisher metric between the two layers.

The same equality holds when s is not one.
There is also an interesting property that
\begin{equation}
M_{n,n+m}^{2}=1-\frac{|\langle V_{\Lambda-n}|V_{\Lambda-n-m}\rangle|^2}{\langle V_{\Lambda-n}|V_{\Lambda-n}\rangle\langle V_{\Lambda-n-m}|V_{\Lambda-n-m}\rangle}=(p^{m-m\Delta}-1)^{2}p^{2n-2n\Delta}\lambda^2
\end{equation}
in which \(M_{n,n+m}\) represents the square root of fisher metric between the nth layer and (n+m)th layer, and
\begin{equation}
M_{ij}+M_{jk}=M_{ik}
\end{equation}

\bibliographystyle{utphys}
\bibliography{emergentE}

\end{document}